\documentclass[12pt,preprint]{aastex}
\usepackage{epsfig}

\received{2003 May 19}
\begin{document}

\title{Collapse of Magnetized Singular Isothermal Toroids: I. Non-Rotating
Case}

\author{Anthony Allen}
\affil{Institute of Astronomy and Astrophysics, Academia Sinica,
PO BOX 23-141, Taipei 106, Taiwan, R.O.C.} 

\author{Frank H. Shu}
\affil{National Tsing Hua University, 101, Section 2 Kuang Fu Road, 
Hsinchu, Taiwan 300, R.O.C.}

\author{Zhi-Yun Li}
\affil{Department of Astronomy, University of Virginia, Charlottesville, 
VA 22903}

\begin{abstract}
We study numerically the collapse of non-rotating, self-gravitating, 
magnetized, singular isothermal toroids characterized by sound speed, $a$, and 
level of magnetic to thermal support, $H_0$. 
In qualitative agreement with previous treatments by Galli \& Shu
and other workers, we find that the infalling material 
is deflected by the field lines towards the 
equatorial plane, creating a high-density, flattened structure -- a 
pseudodisk.  The pseudodisk contracts dynamically in the radial direction, 
dragging the field lines threading it into a highly pinched configuration
that resembles a split monopole.
The oppositely directed field lines across the midplane and the large implied
stresses may play a role in how magnetic flux is lost in the actual situation
in the presence of finite resistivity or ambipolar diffusion.
The infall rate into the central regions is given to 5\% uncertainty by the 
formula, $\dot M = (1+H_0)a^3/G$, where $G$ is the universal gravitational constant,
anticipated by semi-analytical studies of
the self-similar gravitational collapses of the
singular isothermal sphere and isopedically
magnetized disks.   The introduction of finite initial rotation results in a
complex interplay between pseudodisk and true (Keplerian) disk formation
that is examined in a companion paper. 
\end{abstract}


\section{Introduction}

\subsection{Low-Mass Star-Formation Theory}

Stars form in the dense 
cores of molecular clouds (Myers 1995, Evans 1999).
The nearby cores that form low-mass stars have a typical diameter
$\sim$ 0.1 pc, a number density $\sim 3\times 10^4$ cm$^{-3}$, a mass 
ranging from a fraction of a solar mass to about 10 $M_\odot$,
and an axial ratio for flattening of
typically 2:1.  Agents
other than isotropic thermal or turbulent pressures must help to support cores
against their self-gravity, although there is some debate
whether the true shapes are oblate, prolate, or triaxial
(Myers 1998, personal communication; see also Jones \& Basu 2002). 
Observed cloud rotation rates are generally too small to account
for the observed flattening (Goodman et al. 1993), and tidal effects from
neighboring cores cannot explain the flattening of inner density contours.
This leaves magnetic fields, implicated in many other physical processes of
importance in contemporary star formation (see the reviews of
Shu et al. 1987, 1999).

In one scenario, ambipolar diffusion 
leads to a continued contraction of a cloud core with ever growing
central concentration (Nakano 1979, Lizano \& Shu 1989, Basu \& Mouschovias
1994).  In a finite time given by slippage of neutrals past ions and magnetic fields,
the molecular cloud core reaches a configuration
where the central density formally tries to reach infinite values.
Shu (1995) proposed the name ``gravomagneto catastrophe'' for this mechanism by analogy
with the ``gravothermal
catastrophe'' (Lynden-Bell \& Wood 1968, Lynden-Bell \& Eggleton 1980,
Cohn 1980) that overtakes globular-cluster core-evolution
because of the diffusion of random velocities by stellar
encounters.

Myers (1999) has emphasized that inward
fluid velocities of magnitude $\sim 0.1$ km s$^{-1}$
exist at large radii in some starless cores,
and that this feature is not predicted by any of the ambipolar-diffusion
calculations cited above (see, however, Li 1998a and Ciolek \& Basu 2000).  
Myers \& Lazarian (1998) suggest instead that
pre-existing inward motions prior to true gravitational
collapse and star formation are better modeled as resulting
from the decay of turbulence.  If the sequence of decaying
turbulent parameter $K = 12, 10, 6, 2$ in Figure 2 of Lizano \& Shu (1989)
were to occur over $\sim$ a few Myr, then inward motions of
order half the isothermal sound speed, $0.5 a\sim$ 0.1 km s$^{-1}$,
could indeed be generated at distances of
order $0.1$ pc from the center of the cloud core.  Large-scale inward motions and
generally shorter time scales may distinguish the
``gravoturbo catastrophe'' resulting from the decay of
turbulence in a pre-existing, magnetically supercritical, core
with dimensionless mass-to-flux ratio $\lambda > 1$ (see Li \& Shu 1966
for the definition of $\lambda$) from a core that
reaches supercriticality via ambipolar diffusion.  As we shall see later,
inclusion of initial inward motions of order $0.5 \, a$ increases the
net central mass-accretion rate due to subsequent gravitational collapse
by about a factor of 2.

\subsection{The Pivotal State}

Li \& Shu (1996) termed the cloud configuration when the central 
isothermal concentration first becomes formally infinite
as the ``pivotal'' state ($t \equiv 0$).  This state
separates the nearly quasi-static phase of core evolution ($t < 0$)
that leads to gravomagneto or gravoturbo catastrophe ($t=0$)
from the fully dynamic phase of protostellar accretion ($t > 0$).
(See the first two stages depicted in Fig. 7 of Shu, Adams,
\& Lizano 1987.)  The numerical simulations indicate
that the pivotal states have a number of simplifying properties,
which motivated Li \& Shu to approximate them as scale-free
{\it equilibria} with the power-law radial dependences for the density
and magnetic flux function:
\begin{equation}
\rho(r,\theta)={a^2\over 2\pi G r^2}R(\theta), \qquad
\Phi(r,\theta)={4\pi a^2 r\over G^{1/2}}\phi(\theta).
\label{rhoPhi}
\end{equation}
In equation (\ref{rhoPhi}), we have adopted
a spherical polar coordinate system $(r,\theta,\varphi)$,
and $a$ is the isothermal sound speed of the cloud, while $R(\theta)$ and 
$\phi(\theta)$ are the dimensionless angular distribution functions for
the density and magnetic flux given by force balance along and
across field lines.  The resulting 
differential equations and boundary conditions yield a linear
sequence of possible solutions, characterized by a single
dimensionless free parameter, $H_0$, which represents
the fractional over-density supported by the magnetic field
above that supported by thermal pressure.
Comparison with the typical degree of elongation
of observed cores suggests that the 
over-density factor $H_0\approx 0.5 - 1$ (corresponding to
$\lambda \approx 2$).

\begin{figure}
  \begin{center}
    \plotone{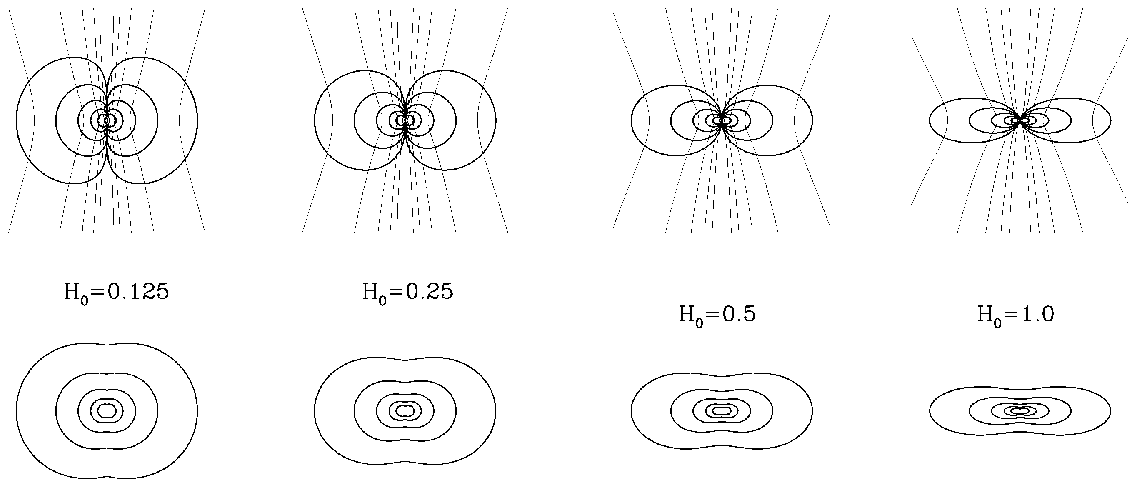}
\caption{Top: Isodensity contours and magnetic field lines for a few 
selected values of $H_0$.  
Bottom: Equatorial plane projection of the 
column density (taken from Shu et al. 1999).}
\label{fig:crete2}
  \end{center}
\end{figure}

In Figure ~\ref{fig:crete2}, we plot isodensity contours and field lines
for the cases $H_0$ = 0.125, 0.25, 0.5, 1.0.  The corresponding values of
$\lambda$ from Table 1 of Li \& Shu (1966) are $\lambda$ =
8.38, 4.51, 2.66, 1.79.
Below each case are the column density contours
that would be seen by an observer in the equatorial
plane of the configuration. 
The presence of very low-density regions
at the poles of the density toroid has profound implications
for the shape and kinematics of bipolar molecular outflows (Li \& Shu 1996,
see also Torrelles {et al.} 1983, 1994).

\subsection{Self-Similar Collapse of Singular Isothermal Toroids}

It is well-known that the collapse of the singular isothermal
sphere (corresponding to the case $H_0 = 0$ in the previous
section) occurs in a {\it self-similar} manner (Shu 1977).
The same holds true for any member in the sequence of
isopedically magnetized, singular isothermal toroids (SITs), whether we
start them from rest, or with a initial, radially directed, velocity field that is
a uniform fraction of the isothermal sound speed $a$.
For given $H_0$, we have only two dimensional
parameters in the problem, $a$ and $G$.  From $a$ and $G$,
we cannot form dimensions for all required physical quantities without
using the independent variables of the problem themselves:
$r$, $\theta$, and $t$.  
The quantities $r$ and $at$ both have the dimensions of length,
and they can be combined to
form the dimensionless similarity variable
\begin{equation}
\label{xdef}
x=r/at ,
\end{equation}
which can then appear, along with $\theta$,
as the argument for nondimensional
mathematical functions.  Thus, the time-dependent axisymmetric solution that
describes the $t > 0$ evolution of the unstable
equilibrium state (\ref{rhoPhi}) must take
the {\it self-similar} form:
\begin{equation}
\rho (r, \theta, t) = {\alpha (x,\theta)\over 4\pi Gt^2},\;\;
{\bf u}(r,\theta, t) = a {\bf v}(x,\theta),\;\;
{\bf B}(r,\theta,t) = {a\over G^{1/2}t}{\bf b}(x,\theta),
\label{rhouB}
\end{equation}
where $\bf u$ and $\bf B$ are the poloidal fluid velocity
and magnetic field, respectively.
The dimensionless dependent variables
$\alpha (x,\theta)$, ${\bf v}(x,\theta)$, and ${\bf b}(x,\theta)$
are called the {\it reduced} density, velocity, and magnetic field.
Solutions with the form (\ref{rhouB}) are called
``self-similar'' because, apart from a simple scaling
of dependent and independent variables, 
the solution at one instant in time looks the same
as the solution at another instant in time.

For the case $H_0$ = 0 (corresponding
to ${\bf b} = 0$),
we may assume that the flow occurs spherically
symmetrically.   With no $\theta$ dependence
and with ${\bf v}$ having only a radial component,
the similarity solution has a simple description if the SIS is started from rest
(Shu 1977).  At $t=0^+$, consider initiating collapse of the densest innermost
portions of a (nonrotating and unmagnetized)
singular isothermal sphere to form
a condensed central object (a protostar, whose physical
dimensions are so small as to allow us here to approximate
it as a point mass).  An observer located in the outer portions
of the cloud does not realize anything has happened gravitationally,
because Newton's theorem for the gravitational field
of spherical objects misleads the observer into thinking
that all the mass interior to the observer's location
equivalently lies at the center anyway.  Only when a
sound wave traveling outwards at the speed $a$ reaches
the observer, does the observer realize, hydrodynamically,
that the bottom has dropped out.  This information 
initiates a wave of falling whose
location at time $t$ has a head described by the locus $r = at$.
The similarity variable description for the same locus
is $x=1$.  For $x$ larger than 1 (or $r>at$), the density
profile retains its unperturbed value $\alpha = 2/x^2$
(or $\rho = a^2/2\pi Gr^2$), and the fluid is at rest,
${\bf v} =0$ (or ${\bf u}=0$).

The solution interior to the head of the expansion wave
is modified from the initial equilibrium.
At $x$ = 1, the solutions for $\alpha$ and $v$
connect without a jump, but with a discontinuous first derivative
(because of the presence of the head of the wave)
to the unperturbed state corresponding to the singular
isothermal sphere.  Near
the origin, the solutions take the
asymptotic forms,
\begin{equation}
\label{innerasymp}
\alpha \rightarrow \left[ {m_0\over 2x^3}\right]^{1/2},
\qquad v(x) \rightarrow -\left[ {2m_0\over x}\right]^{1/2} ,
\qquad {\rm as} \qquad x \rightarrow 0,
\end{equation}
that correspond to gravitational free-fall onto
a (reduced) central mass found by numerical integration
to be $m_0 = 0.975$.
In dimensional units,
the origin contains a central mass point (star) that grows
linearly with time:
\begin{equation}
\label{centralmass}
M = m_0 {a^3t\over G} .
\end{equation}
Equation (\ref{centralmass}) constitutes
the most important property of the solution --
the conclusion that the gravitational collapse of a
singular isothermal sphere yields, not a characteristic
mass, but a characteristic {\it mass infall rate} given by
\begin{equation}
\label{Mdotin}
\dot M = m_0\, {a^3\over G} .
\end{equation}

Inclusion of the effects of finite levels of the magnetic field
(i.e., $H_0 \neq 0)$ does not change the qualitative situation.
The collapse still occurs self-similarly from ``inside-out,''
but the departures from spherical symmetry implies that
non-zero motions can be induced {\it gravitationally} ahead
of the front of an expansion wave (see also Terebey,
Shu, \& Cassen 1984 for the case when the cause of departure
from spherical symmetry is rotation).  If the
gravitational precursor motions are sufficiently compressive, 
the expansion wave, now propagating outward at the fast
MHD speed, may even be preceded by a weak shock wave
(Li \& Shu 1997).  But the main new feature introduced
by the presence of the magnetic field is a flattening
of the flow in the inner regions to form a ``pseudodisk''
(see the singular perturbational treatment by Galli \& Shu 1993a,b
of this problem). We will concentrate on the formation and 
structure of this pseudodisk in our numerical calculations, and we shall
focus for simplicity most of our attention on equilibrium configurations
started from rest.  At the end, however, we will discuss the modifications
introduced by starting with uniform inward motions present already at the pivotal instant.

Strict field freezing implies that the mass-to-flux ratio is a
conserved quantity: $\lambda$ = const.  Except for the time-dependent behavior
that tapers the surface density distribution near the edge,
pseudodisks formed from the gravitational collapse of magnetized SITs
therefore will resemble nonrotating, magnetized,
singular isothermal disks 
with values of the dimensionless mass-to-flux
ratio $\lambda$ greater than, but close to, unity. 
Since such disks, started from rest, collapse to yield a central mass-accretion
rate $\dot M = m_0 a^3/G$ with $m_0 = 1.05(1+H_0)$, Li \& Shu (1997) 
speculated
that magnetized SITs, for arbitrary initial values of $H_0$
from 0 to $\infty$, would develop central mass-infall rates
well-approximated (to 5\% accuracy) by the simple formula:
\begin{equation}
\label{magMdotin}
\dot M = (1+H_0){a^3\over G} ,
\end{equation}
where $H_0$ is the degree of support by magnetic fields provided in the
pivotal configuration.
One of the major goals of the present numerical simulation is to justify explicitly
the above equation.

For $a \approx 0.2$ km s$^{-1}$ and $H_0 \approx 0.5$ (Taurus region), 
equation
(\ref{magMdotin}) yields $\dot M \approx 3\times
10^{-6}$ $M_\odot$ yr$^{-1}$, in agreement with the
spectral energy distribution (SED) and light-scattering models of
Kenyon et al. (1993a, b).  The time $t$ required
to build the typical T Tauri star of mass 0.5 $M_\odot$
is then $1.7\times 10^5$ yr.  The inference of constant accretion rates,
with associated mass-accumulation time scales of $\sim 10^5$ yr and observed
luminosities according to standard models of protostellar evolution
(e.g. Stahler et al. 1980, Palla \& Stahler 1990), is consistent
with the survey of Hirano et al. (2002) of 17 protostellar sources in
isolated dark clouds, Taurus, and Rho Ophiuchus star-forming regions.
They are not consistent with claims for coefficients in the relationship,
$\dot M \propto a^3/G$,  which are either one to two orders of magnitude
larger or highly time-variable, that have been claimed 
in the literature based on numerical simulations 
(for low-mass star-forming regions) starting from conditions rather 
different from singular isothermal
toroids (see, e.g., Tomisaka 2002, who started the 
collapse calculation from an infinitely long cylinder, with the 
collapse induced by a sinusoidal perturbation in density along 
its length). 
Myers \& Fuller (1993; see also Myers, Ladd, \& Fuller 1991; Myers 1995)
made the intriguing suggestion
that total mass-accumulation time scales $\sim 10^5$ yr may apply across
the board, from the formation of the lowest-mass to the highest-mass stars.
If this suggestion has merit, and if we continue to use {\it static},
magnetized SITs to model the pivotal states of such objects,
then the final mass of the formed star must scale with the
factor $a^3(1+H_0)$, with this factor
being larger to form more massive stars.  In other words,
massive star-forming regions must be hotter, or more highly magnetized,
or more turbulent (if we heuristically include the effect of turbulent
support into $H_0$), than low-mass star-forming regions (see the reviews
of Shu et al. 1987 and Evans 1999).

Non-isothermal equations of state
to mimic the effects of turbulence (where the signal speeds are larger
at lower densities) can yield
pivotal states with density laws that are shallower than
the $\rho \propto r^{-2}$ dependences that are considered in this paper
(see, e.g., Lizano \& Shu 1989, McLaughlin and Pudritz 1997,
Galli et al. 1999).  
The gravitational collapse
of such non-isothermal configurations will
yield mass infall rates $\dot M$ that increase
with time. In contrast, pre-existing inward motions in the central 
region at the pivotal instant $t=0$ tend to yield $\dot M$ that 
decreases with time (Basu 1997, Ciolek \& K\"onigl
1998, Li 1998a).  When added to the support provided quasistatically
by thermal pressure, all of these mechanisms
{\it enhance} the mass infall rate above the base value
given by the inside-out collapse solution for the
initially static, singular isothermal sphere: $\dot M = 0.975 a^3/G$.
None of the considerations imply
that $\dot M$ should {\it cut off} after some definite time
and leave us with a star with a well-defined mass.
We return to this difficulty in the last section of this paper
(see also the discussion in Shu \& Terebey 1984, Shu et al. 1999).

In any case, whether or not ideally magnetized SITs
represent good pivotal states at $t=0$
for star formation models, the self-similar nature of their $t > 0$ behavior
greatly simplify both the collapse calculations and data storage requirements.
For the latter, we need to store the reduced variables that are only
functions of the similarity coordinates $(x,\theta)$.  Moreover, these reduced
variables depend only on a single dimensionless parameter $H_0$
of the initial state (deducible under the assumption of field freezing
from a measurement of $\lambda$ at any instant during the dynamical collapse).
Comparisons with the variables of actual
observations at any dimensional time $t$ can then be obtained from
the single parameter family of solutions by appropriate scalings of
dependent and independent variables by various factors of $t$, $a$, and $G$.

\section{Numerical Method and Simulation Setup}

We follow the inside-out collapse of magnetized singular isothermal toroids 
using Zeus2D (Stone and Norman, 1992a,b), 
which solves the ideal MHD equations 
in 3 dimensions with an axis of symmetry. Ideal MHD is a reasonable
approximation during the dynamic collapse, since there is little time 
for the magnetic field to diffuse outwards relative to the bulk matter
(see, e.g., Galli \& Shu 1993a, b). We 
are interested in the dynamics of the collapsing flow not too close to
the center, where IR radiation can escape freely and an isothermal 
equation of state can be adopted. We note that violation of the isothermal
assumption could affect the vertical 
structure of pseudodisks, which may be bounded by accretion shocks on 
their upper and lower surfaces. 

Zeus2D is a well tested, general purpose, publicly available package. 
Applying it to our specific problem was not without difficulties. 
These difficulties and the modifications we developed to overcome them
are discussed below.

\subsection{Polar Axis and Central Cell Modifications}

The family of models studied in this paper suffers from several numerically 
inconvenient features, including a point mass at the origin that grows 
linearly with time and a near vacuum in the polar region. The near-vacuum 
region is further complicated by finite magnetic field strengths that 
imply nearly infinite Alfv\'{e}n speeds.  Close to the polar axis, field 
lines run nearly parallel to the polar axis so that there is little 
(vertical) magnetic support of fluid.  With an increasing point mass 
at the origin, velocities grow and densities deplete with time. Small 
irregular Lorentz forces induced by a wave traveling up a field line 
can produce artificially large fluid velocities in the near vacuum that 
cause the cell to deplete faster than gravity can actually draw down 
the gas.  

We allow mass to accumulate in the central cell, and modify the ZEUS 
code so that only equations dealing with magnetic fields see the 
matter in the central cell\footnote{If the magnetic field isn't tied 
to the central mass, it will slip outward, resulting in artificial 
diffusion.}.  Particularly, the large pressure in the central cell is
ignored in computing pressure gradients in the force equation, as in 
the standard sink-cell method. The central cell itself does not appear to 
cause any numerical problem. Those close to the origin along the 
polar axis are, however, vulnerable to instability. Usually, one cell 
would become depleted 
of fluid so that either the Alfv\'{e}n speed would become too large (in 
strong field cases) or fluid velocities would become infinite (weak field 
cases).  The instability is purely numerical; a vacuum moving at infinite
speeds is physically meaningless.
The representation of a spherical singularity
by a cylindrical central cell may aggravate the difficulties,
eventually halting the simulation.  In some cases, 
we could mitigate this instability 
by imposing the monopole solution for velocity (Galli 
\& Shu, 1993a,b) along the polar axis region where $x<0.1$.  
The corrections change only the features of the near vacuum region 
around the polar axis and have no influence on other regions of 
physical interest in the simulation.  

\subsection{Alfv\'{e}n Limiter in Low Density Polar Region} 
\label{alflim}

For the magnetized singular isothermal toroids, which are the initial 
configurations of our simulations, the magnetic field close to the polar 
axis scales as $B \propto 1/r$. The density close to the polar axis 
scales as $\rho \propto \theta^{4H_0}/r^2$. Thus, the Alfv\'{e}n 
speed near the polar axis, $v_a \propto |B|/\sqrt{\rho} \propto 
\theta^{-2H_0}$, becomes increasingly divergent as $H_0$ increases.  
The maximum timestep in an explicit code such as Zeus2D is limited by 
the Courant 
condition, which sets the maximum timestep to be of order the fastest 
cell-crossing time of Alfv{\'e}n wave for all the cells.  Thus, as $H_0$ 
increases, the maximum timestep $\Delta t \propto \theta^{2H_0}$ becomes 
increasingly small for cells close to the polar axis, creating a problem
for the simulations.  

We circumvent the Alfv{\'e}n problem by artificially decreasing the 
Alfv\'{e}n speed when the speed becomes too high. This fast Alfv\'{e}n 
impediment to MHD simulations is 
not new.  One method to overcome the problem that has 
had some success (Miller \& Stone, 2000), involves inclusion of the 
displacement current in Maxwell's induction equation, but with a 
reduced value for the speed of light (Boris, 1970).  This sets an upper bound
to the Alfv\'{e}n speed.  We note that the use of the Alfv\'en limiter
does not destroy the self-similarity of the problem as long as we set
the speed of light equal to a constant multiple of the speed of sound
$a$.  Of course, there is a price to pay for the larger 
timestep possible; one can not expect accurate results in regions where the 
Alfv\'{e}n speed has been truncated (e.g. near the polar axis).  This 
price is worth paying, as demonstrated below. 

\subsection{Boundary Conditions and Grid}

Aside from the Courant condition, a more subtle issue with the boundary 
conditions required the Alfv\'{e}n limiter to be imposed for our
simulations.
Outer boundary conditions were set either to be divergence-free outflow
(important for the rotating cases to be discussed in our companion paper) or 
calculated from the fluid variables at an earlier time $\tau$ and smaller 
radius within the computational domain by use of the self-similarity of 
the problem:
\begin{equation}
\label{eq:ss}
\rho(r,\theta,t) = 
\left(\frac{\tau}{t}\right)^2\rho\left(r\frac{\tau}{t},\theta,\tau\right), \;\;
{\bf u}(r,\theta,t) = {\bf u}\left(r\frac{\tau}{t},\theta,\tau\right), \;\;
{\bf B}(r,\theta,t) = \frac{\tau}{t}{\bf B} \left(x\frac{\tau}{t},\theta,\tau\right).
\end{equation}
Inner boundary conditions on the polar axis were set to rotational symmetry.
For most runs, there was an inner boundary in the equatorial plane set to 
be reflective symmetry with continuous $B_z$.  However, some tests were 
carried out without the equatorial boundary, producing equivalent
results.  The central cell presents a special problem that we shall return 
to later.

Both sets of outer boundary conditions 
give good results \emph{as long as the outer boundary is far away from the 
regions of interest}.  Strictly speaking, we should require the outer 
boundary to be farther than the fastest signal can travel from the origin 
during the timescale of interest.  Since the speed is infinite along the 
polar axis, it is safer to impose the Alfv\'{e}n limiter in all 
simulations; even though there is no appreciable mass in this region, 
reflections of near-vacuum field could still affect the simulations, 
although this effect did not appear in tests until the expansion wave 
of collapse reached the boundary \emph{in regions of modest density}. In
practice, we find it sufficient to set the outer boundary at several times the
product of the sound speed and time, $at$. 

Unless noted otherwise, we use a logarithmic grid of 200x200 in the 
simulations, with the size of a grid cell increasing by a constant 
factor over that of the adjacent cell interior to it. A factor of 
$1.045$ increase and a reduced light speed of $250 \, a$
gave $5$ orders of magnitude in spatial resolution and $3$ orders 
of magnitude in $at$ resolution. The initial configuration is in a 
mechanical equilibrium to within the errors introduced by discretization.
We induce the inside-out collapse with a small point mass at the center,
whose effect diminishes with time. 

\subsection{Treatment of Self-Gravity} 

The gravitational potential is a mathematical convenience used in 
calculating gravitational
forces.  For an inverse square law density distribution of infinite
extent, the gravitational potential is logarithmically divergent.  For the
purpose of the self-gravity calculation only, density outside the largest
radius within the computational grid is renormalized by subtraction of
$(1+H_0)$ from the density function $R(\theta)$.  This renormalization
removes the divergent monopole term from the gravitational potential 
due to mass outside of the computational domain,
without affecting the forces within the computational grid--
we can add or subtract spherical shells of constant density
without affecting forces \emph{inside} those shells.  

\begin{figure}
  \begin{center}
    \leavevmode
\epsfig{file=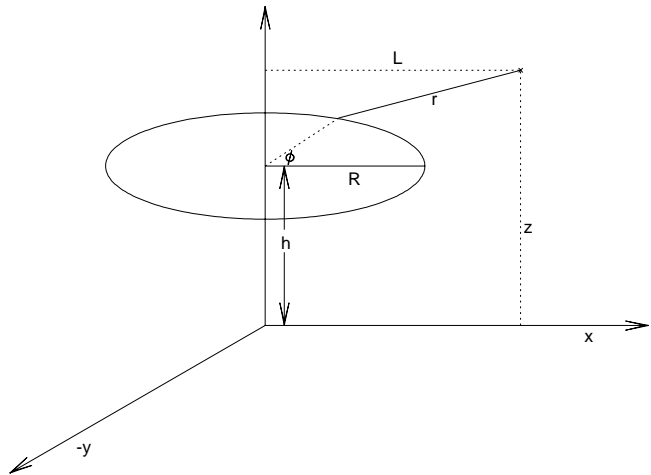,width=4in,clip=,angle=0,bbllx=.5in,bblly=4.5in,bburx=8.5in,bbury=10in}
\caption{Notation for ring integration in the calculation of self-gravity.} 
\label{fig:gravcoord}
\end{center}
\end{figure}

For a cylindrical computational grid, boundary values of the gravitational
potential on the grid's outer boundaries are determined by summation of
cell densities multiplied by a table of geometric factors representing the 
fluid ring for each computational cell and each boundary cell:
\begin{equation}
\mathcal{V}_{\rm ring}= \frac {2G M_{\rm ring} \mathcal{K}\left(\xi\right)}
{\pi \sqrt{(z-h)^2+(R-L)^2}}, \qquad \xi \equiv -\frac{2RL}{(z-h)^2 + (R-L)^2}
\end{equation}
where $\mathcal{K}(\xi)$ is the complete elliptic integral of the first 
kind (see Fig.~\ref{fig:gravcoord} for geometry and notation).  
The gravitational potential due to (renormalized) rings outside the 
computational grid is included out to a radius much larger, typically
$\mathcal{O}(10^4)$, than the length scale of the 
computational grid. Inner boundary values are determined from the 
symmetry of the problem.  After the boundary conditions are specified, 
solution of the Poison equation is left to the sparse matrix solver 
included in the Zeus2D code.  

\section{Numerical Results} 

\subsection{Collapse of $H_0=0.25$ Toroid}

\begin{figure}
    \centering
\leavevmode
        \epsfxsize=.48\textwidth \epsfbox{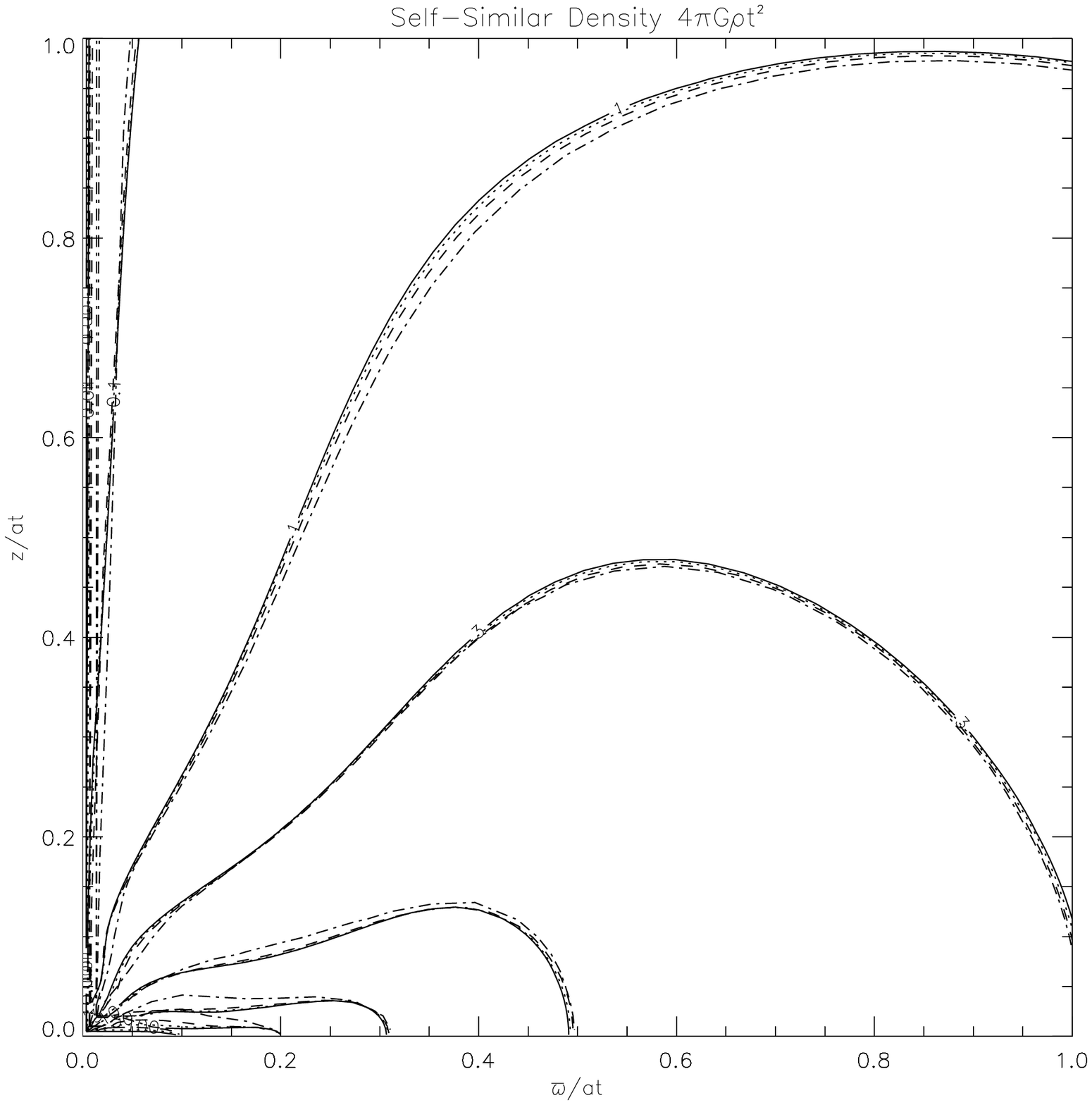} 
        \epsfxsize=.48\textwidth \epsfbox{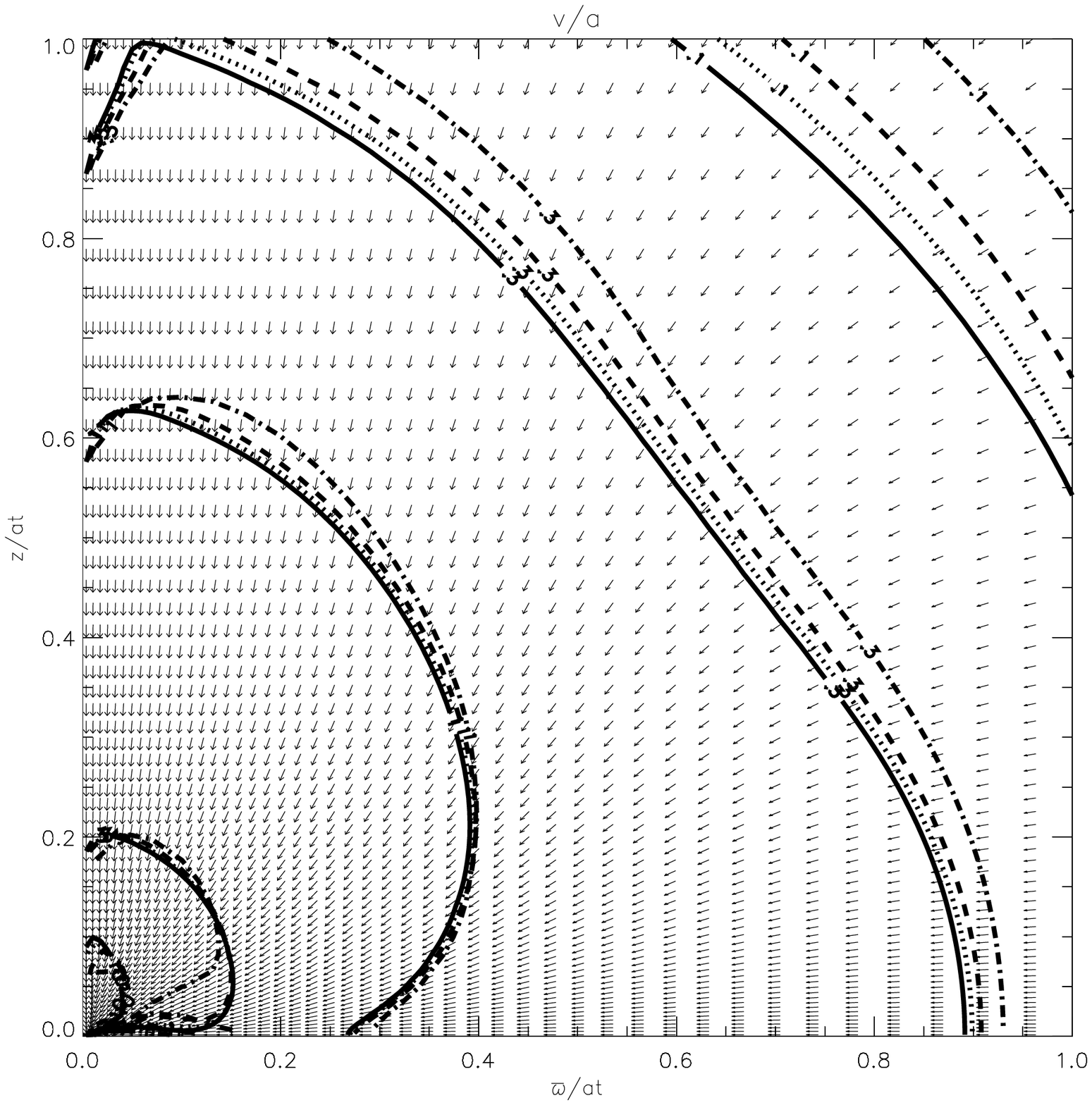} 
        \epsfxsize=.48\textwidth \epsfbox{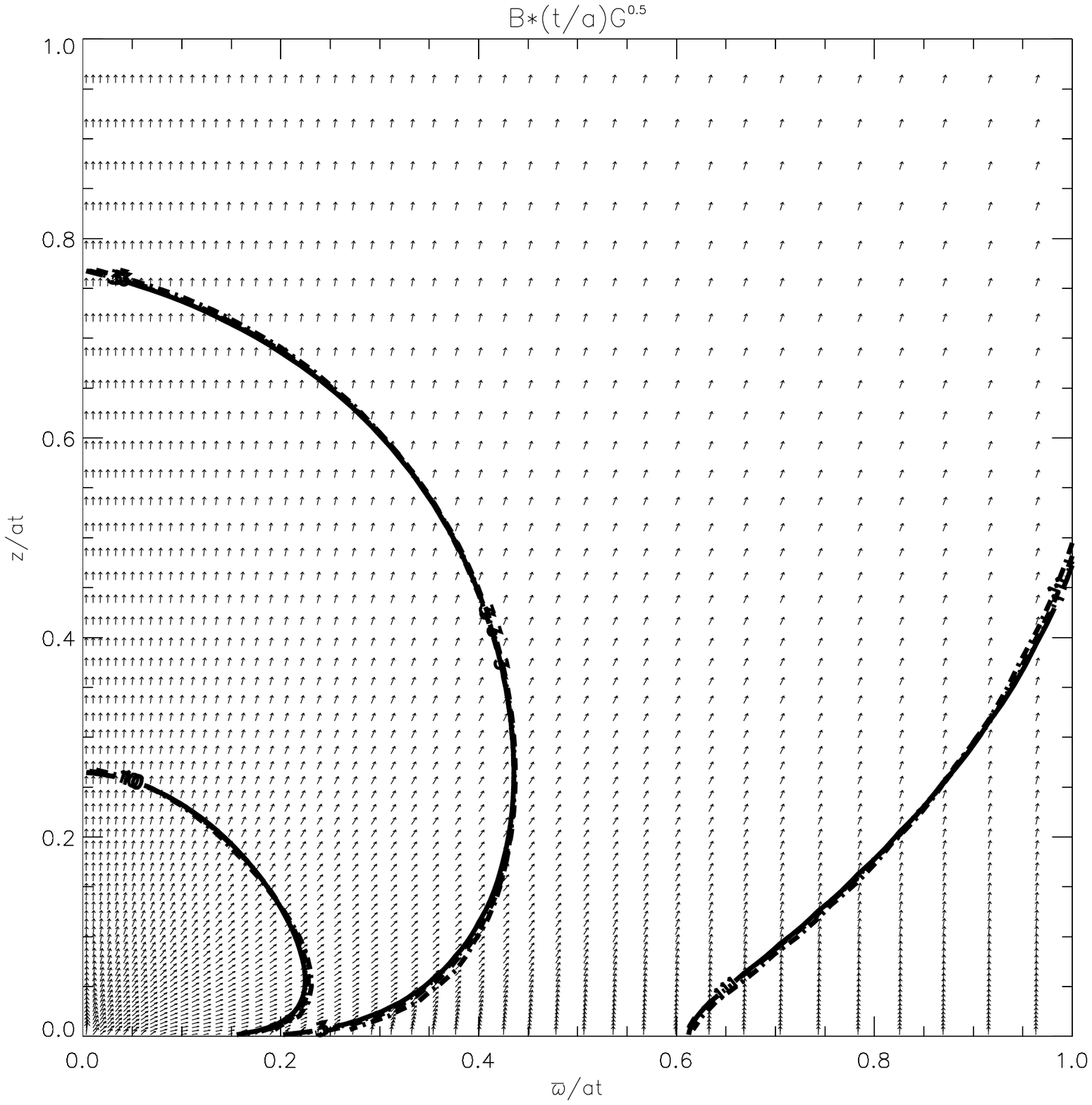} 
        \epsfxsize=.48\textwidth \epsfbox{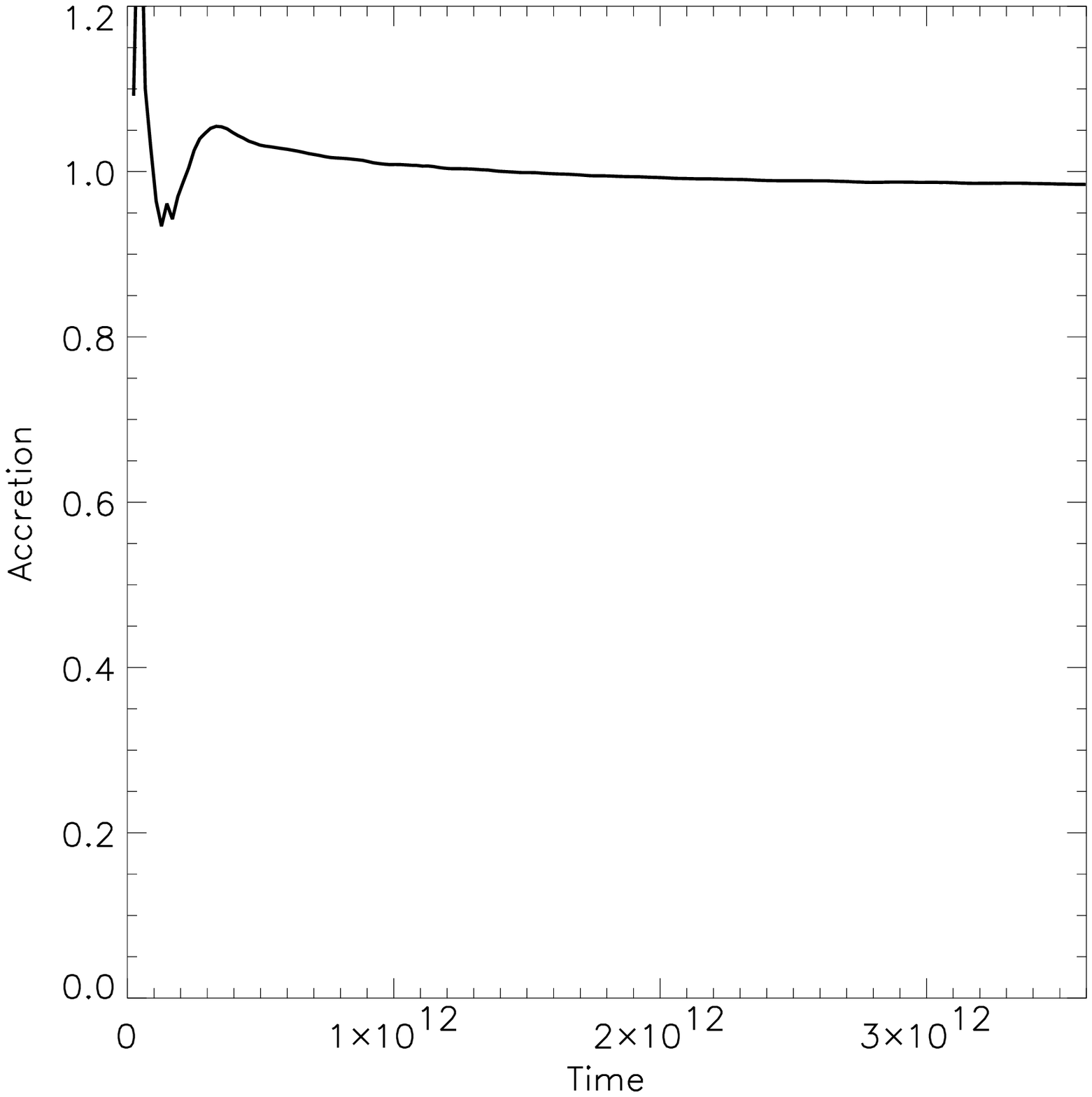}
\caption{Collapse solution of the $H_0=0.25$ case.  Plotted in different 
panels are (a: upper-left) self-similar density at times $0.9\times10^{12}$, $1.8\times 
10^{12}$, $2.7\times 10^{12},$ and $3.6\times 10^{12}$~s as 
dash-dotted, dashed, 
dotted, and solid lines, respectively;  (b: upper-right) self-similar velocity 
in like manner with unit vectors shown only at time $3.6\times 10^{12}s$;
(c: lower-left) self-similar magnetic field in like manner, and (d: lower-right) the accretion 
constant $m_0=G{\dot M}/(1+H_0)a^3$ as a function of time 
in seconds, which asymptotes to 
a number consistent with 0.975. }
\label{fig:n1ss}
\end{figure}

We first consider a fiducial case with $H_0=0.25$ ($\lambda = 4.51$)
to illustrate basic features of the collapse solutions. To induce 
the initial equilibrium toroid to collapse, we add a small pointmass gravitational force at the origin. The added mass 
becomes smaller and smaller compared with the total mass accumulated
in the central cell as the accretion proceeds. As mentioned earlier, the
computations are carried out in the usual time-dependent way, without 
making use of the similarity assumption.  The results at time intervals 
of $\Delta t = 0.9\times 10^{12}$~s 
are displayed in Fig.~\ref{fig:n1ss}, as contour 
plots in the meridional plane of the similarity coordinates $\varpi/at$ 
and $z/at$.  If the numerical calculations were infinitely precise, the
contours for the reduced density $\alpha$, speed $|{\bf v}|$, and 
magnetic field strength $|{\bf b}|$ at different times would all lie 
on top of one another, defining the exact similarity solution.  
Because of the introduction of a small point mass to induce collapse and
errors associated with finite differencing, the contours at different 
times do {\it not} lie exactly atop one another, with the effects of
the initial point mass larger at earlier times and the errors getting
smaller at later times as the nontrivial part of the flow is resolved
by more and more grid points. This behavior is particularly clear in 
Fig.~4d, where the mass accretion rate, defined as the 
numerical derivative of the mass in the central cell with respect 
to time, is plotted.   
Thus, unlike most numerical simulations 
which degrade with the passage of time, these simulations get better 
with time, and the computed values at $t = 3.6\times 10^{12}$ s (the only 
case for which unit vectors showing the directions of ${\bf v}$ and 
${\bf b}$ are displayed) have essentially converged on the correct 
self-similar values.

\begin{figure}
    \centering
    \leavevmode
        \epsfxsize=.48\textwidth \epsfbox{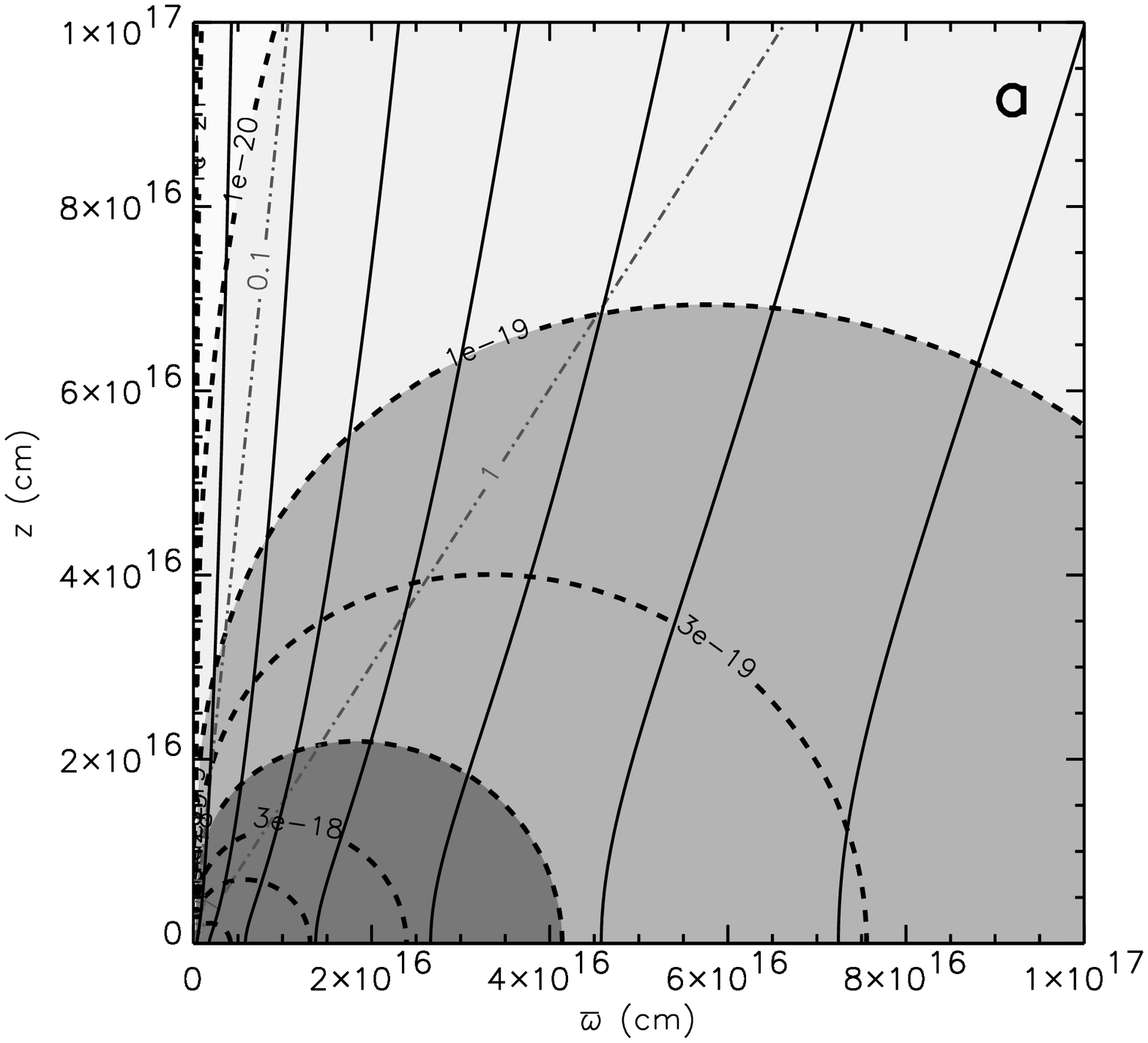} 
        \epsfxsize=.48\textwidth \epsfbox{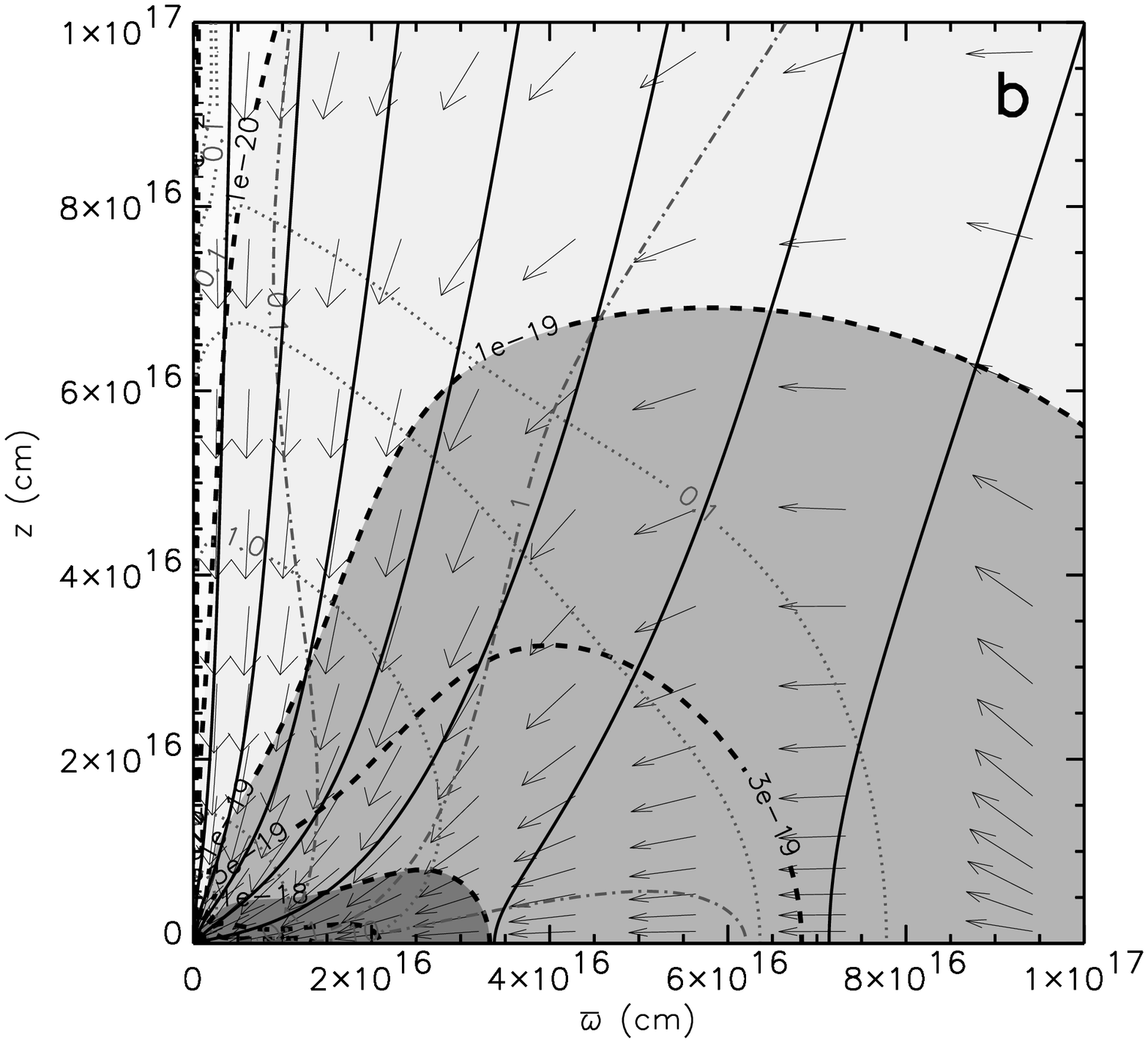} 
        \epsfxsize=.48\textwidth \epsfbox{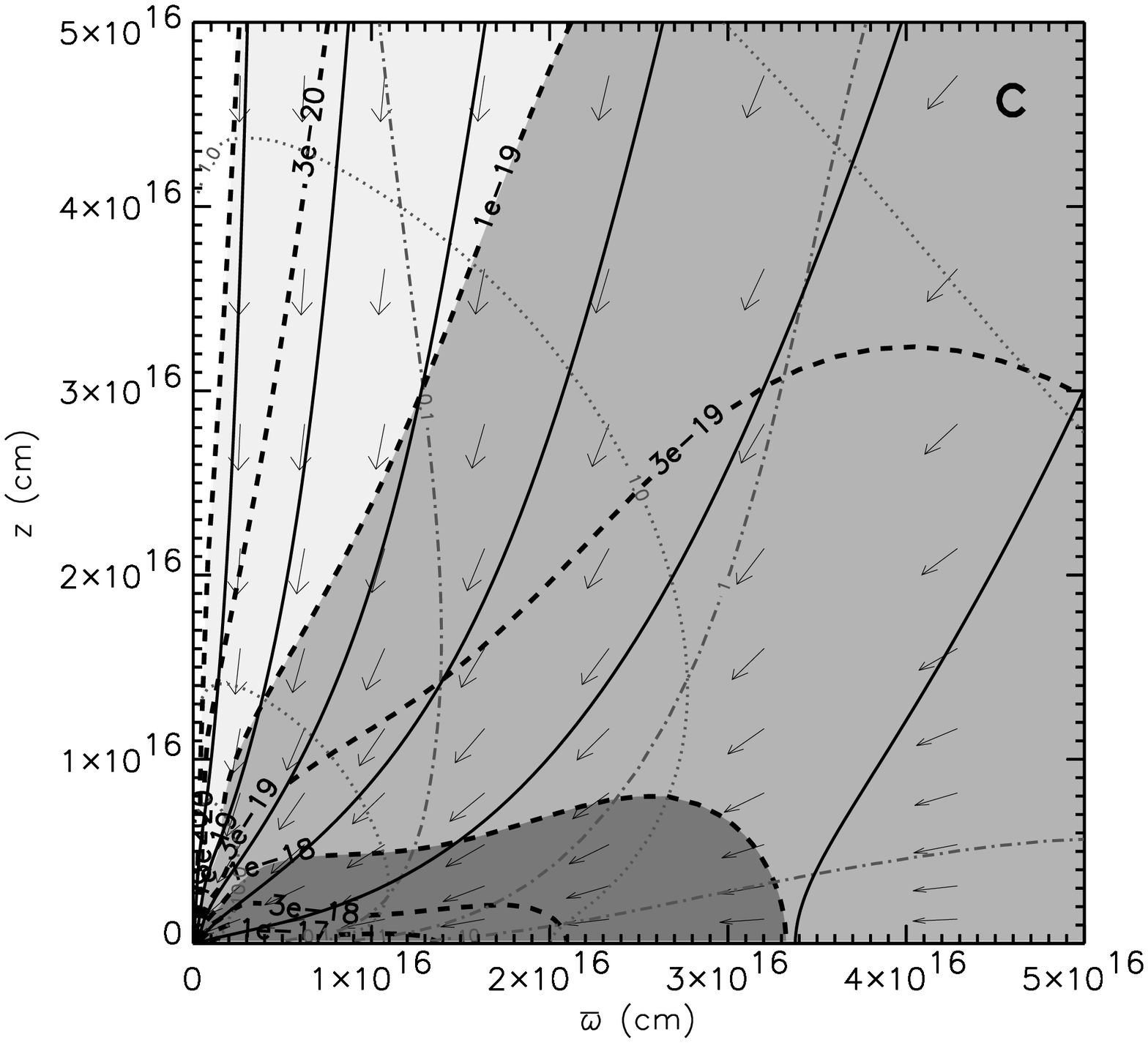} 
        \epsfxsize=.48\textwidth \epsfbox{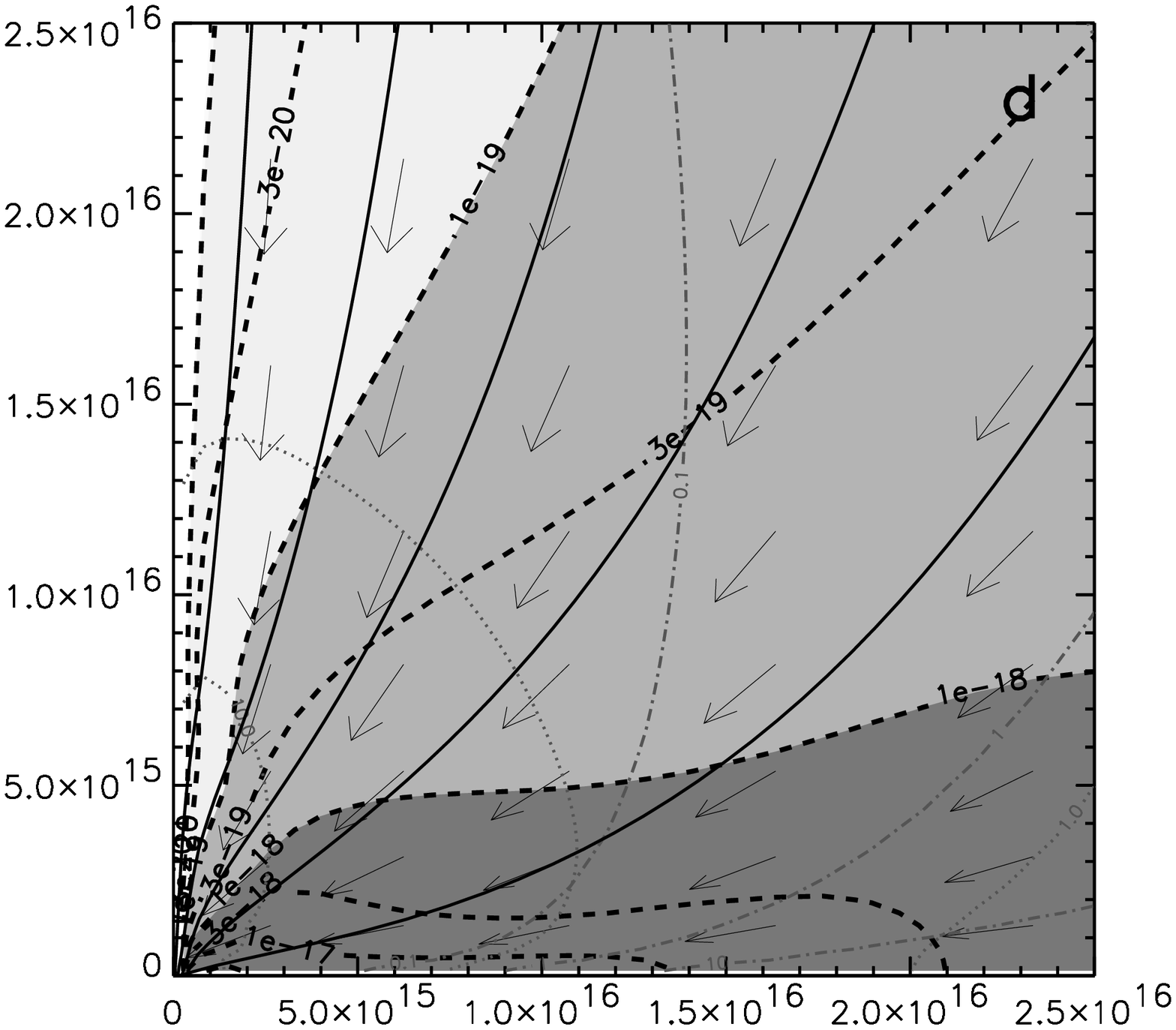}
\caption{Collapse solution for the $H_0=0.25$ case at the initial time $t=0$ 
(a) and the time $t=3.6\times10^{12}$~s (b) shown in cgs units
when we adopt $a=0.2$ km s$^{-1}$. Panels (c) and (d) are closeup 
views of the high-density pseudodisk shown in panel (b). The isodensity 
contours are plotted as dashed lines, with the shades highlighting the
high density regions. The medium-gray contours are bounded by $\rho 
= 10^{-19}$ and $\rho=10^{-18}$~g~cm$^{-3}$, with every second dashed 
contour denoting one order of magnitude in density.  
The magnetic field lines are plotted as solid lines,  
with contours of constant $\beta$ (dash-dotted) superposed. The
velocity is shown by unit vectors, with its magnitude in units 
of the sound speed given by the
dotted contours. }
\label{fig:n1real}
\end{figure}

After the solution has converged in the self-similar space, the solution in 
real coordinates at any time can be easily calculated by a simple scaling, 
as shown in Fig.~\ref{fig:n1real} for $t=3.6\times 10^{12}$~s or $1.1\times 
10^5$~years after the initiation of collapse. In panel (a), we plot the 
initial pre-collapse solution for comparison. Note that for $H_0=0.25$, 
the initial cloud is dominated by the magnetic field in the polar region
within about $30^\circ$ of the axis, where the gas to magnetic pressure 
ratio, $\beta$, is less than unity. Fluid in this region can slide more
or less freely along the field lines, leading to a highly-flattened 
structure in the collapse solution (see panel [b]), which is the pseudodisk 
first discussed by Galli \& Shu (1993a,b, see also Nakamura, Hawana \&
Nakano 1995 and Tomisaka 2002). Closeup views of the pseudodisk
are shown in panels (c) and (d). Note that the pseudodisk is not a static 
structure; collapsing matter is funneled along the field lines in the 
magnetically dominated polar region towards the equatorial plane to form 
the disk and the fluid in the disk is falling radially towards the point
mass at the center, dragging magnetic field lines along with it. Once the 
inflow stops, perhaps by the expansion wave reaching the edge of the cloud 
core, the pseudodisk will disappear into the central point mass.  

The dragging of magnetic field lines by the infalling disk material towards 
the point mass creates a highly pinched field configuration, with a split 
magnetic monopole formed at the origin. The pinched field lines compress 
the disk matter towards the equatorial plane, further enhancing the disk
flattening. Indeed, the region enclosed by the highest density contour 
in panels (b)-(d) of Fig.~\ref{fig:n1real} may not be resolved vertically, 
because of the severe compression due to magnetic pinching. The 
outward-directing tension force of the pinched field lines helps slow 
down the radial collapse in the disk, which contributes to the disk density 
enhancement. The magnetically retarded pseudodisk may be subject to the
well-known magnetic interchange instability in 3D, an issue reserved for
possible future investigation. 

\begin{figure}
    \centering
    \leavevmode
    \psfig{file=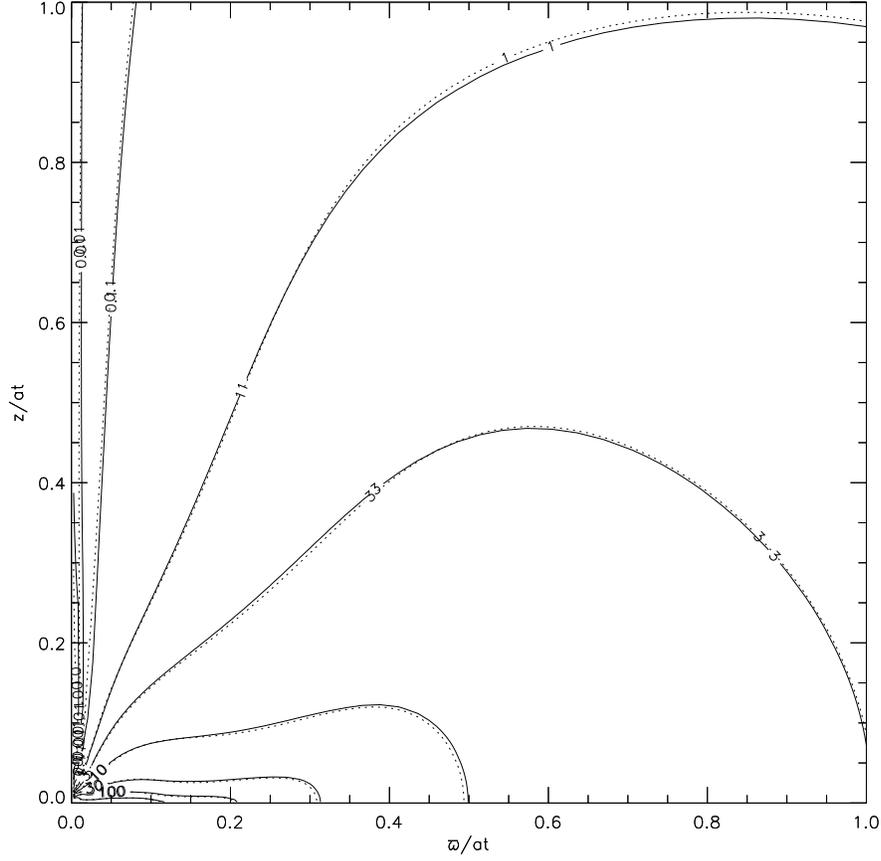,width=0.75\textwidth,angle=0}
\caption{Effects of Alfv{\'e}n limiter on the isodensity contours 
of the collapse solution of the $H_0=0.25$ case shown in similarity 
coordinates.  Solid contour lines 
denote simulations on a 120x120 logarithmic grid with a grid ratio 
of 1.045 and no displacement current limited speed of light.  
Dotted contours denote simulations on a 150x150 logarithmic grid 
with a grid ratio of 1.035 and the speed of light limited to 
$1700 a$ where $a$ is the isothermal 
sound speed.  The effect of the displacement current can only be 
seen in the extremely low density velocity field near the polar axis.  Minor differences from Figure 6c are the result of differing initial point mass perturbations.} 
\label{fig:compare}
\end{figure}

Before discussing cases other than $H_0=0.25$, we note that the application 
of Alfv{\'e}n limiter described in \S~\ref{alflim} does not change the
collapse solution significantly. This is demonstrated in 
Fig.~\ref{fig:compare}
for the $H_0=0.25$, where collapse solution can be obtained with or
without the Alfv{\'e}n limiter. Note that the isodensity contours for
the two solutions are nearly identical, except in a small, near vacuum 
region close to the axis. For higher values of $H_0$, it becomes more
difficult to obtain collapse solution without the aid of the Alfv{\'e}n 
limiter. 

\subsection{Dependence on $H_0$}

\begin{figure}
    \centering
    \leavevmode
        \epsfxsize=.48\textwidth \epsfbox{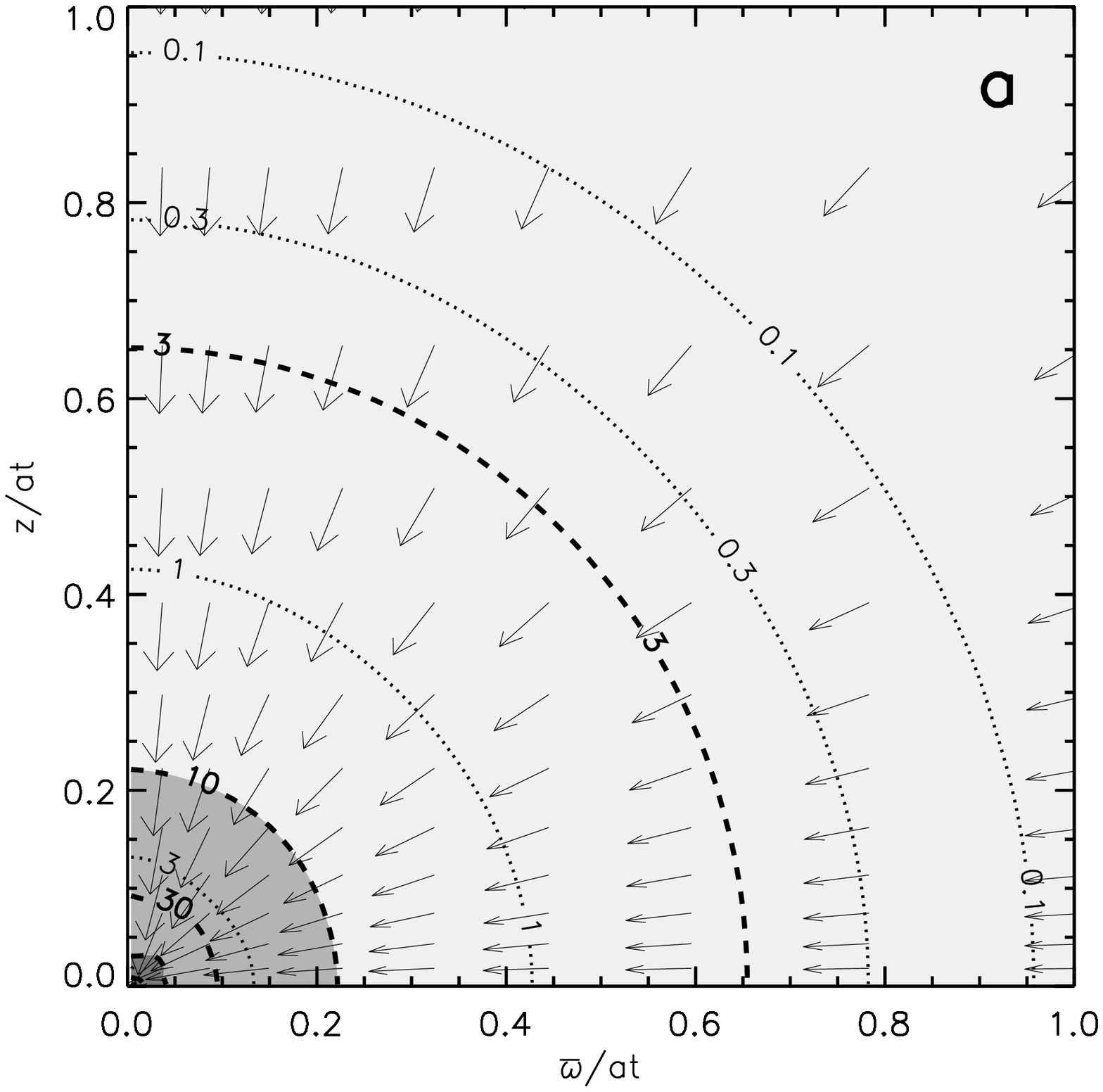} 
        \epsfxsize=.48\textwidth \epsfbox{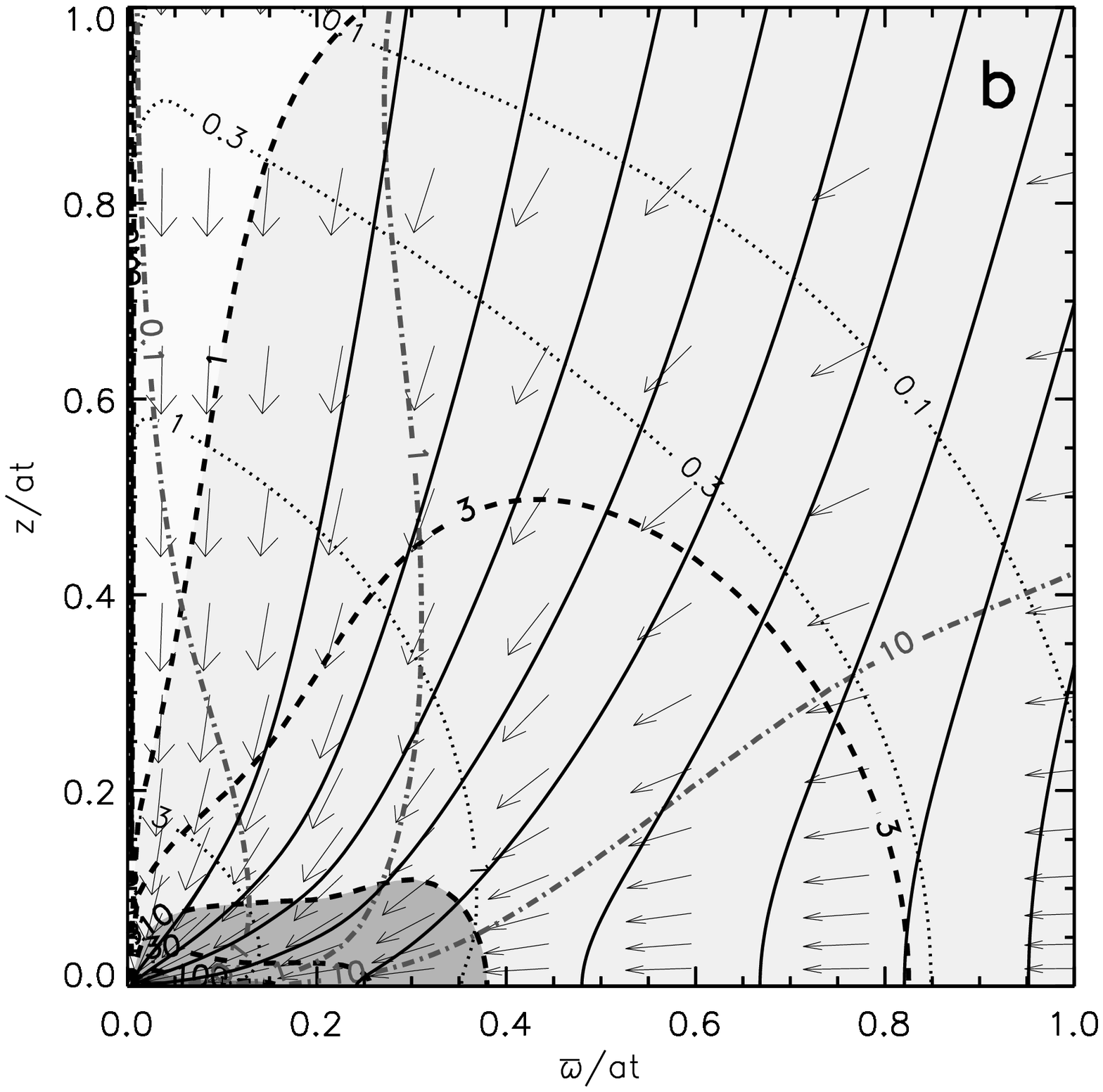} 
        \epsfxsize=.48\textwidth \epsfbox{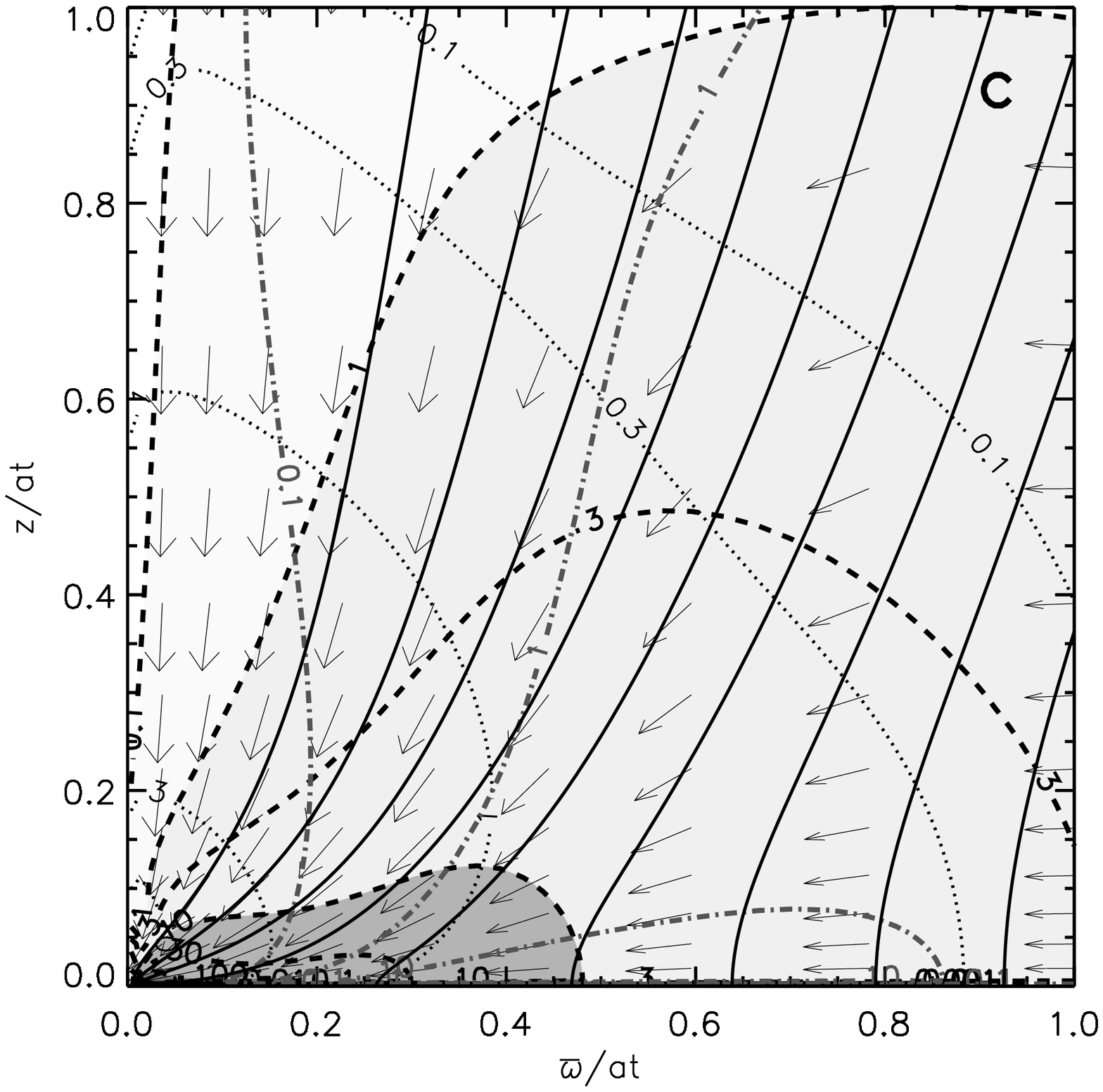} 
        \epsfxsize=.48\textwidth \epsfbox{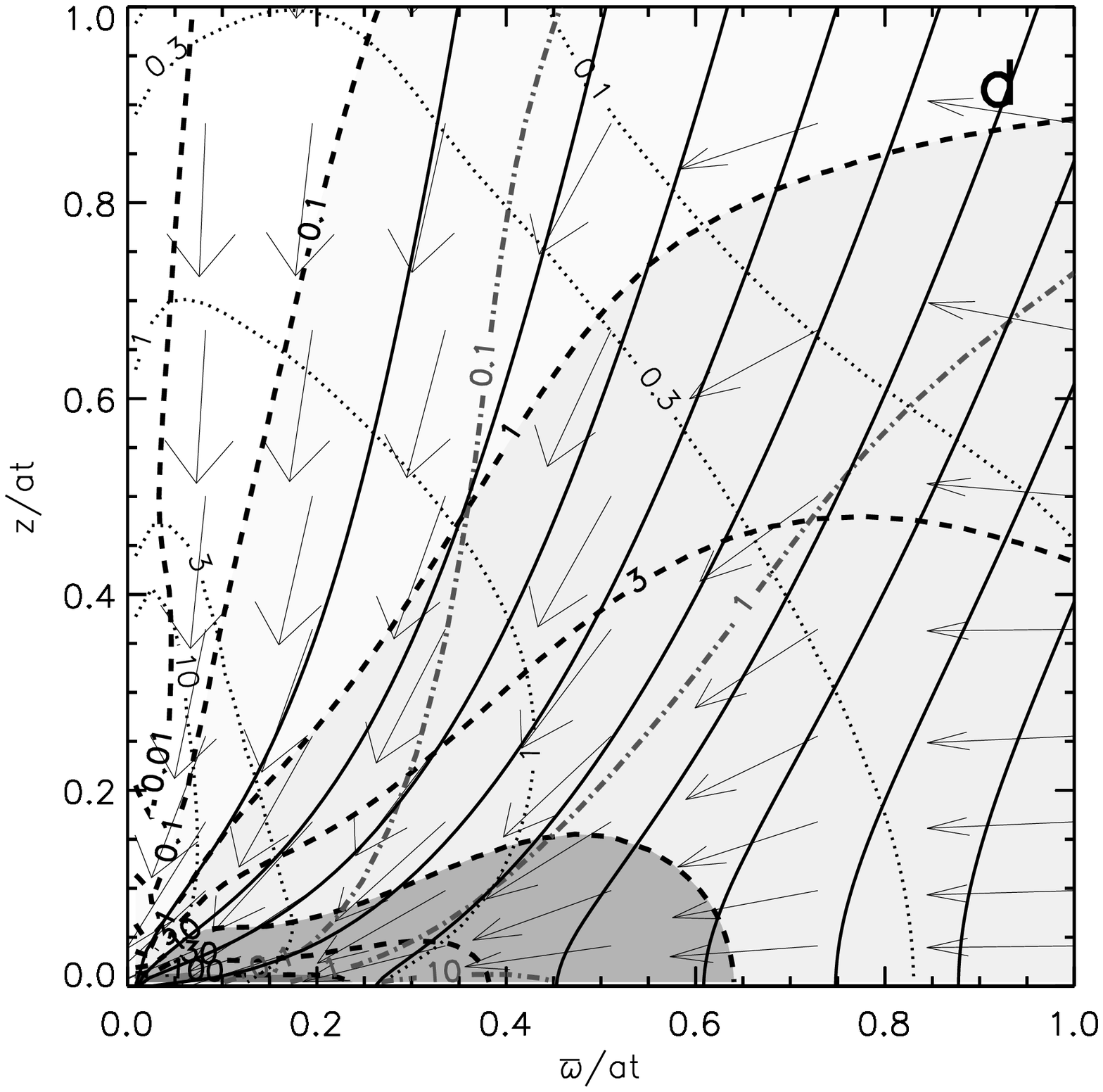}
\caption{Collapse solutions for different degrees of magnetization 
characterized by $H_0=0,0.125,0.25,0.5$. The contours of constant 
self-similar density $4\pi Gt^2\rho$ are plotted as dashed lines, 
with the shades highlighting the high density regions. The magnetic 
field lines are plotted as solid lines,  with contours of constant 
$\beta$ (dash-dot-dashed) superposed. The velocity in every fifth 
cell is shown by unit vectors, with its magnitude in units of the 
sound speed given by the dotted contours.  Field lines are not the 
same across all figures; examine the $\beta$ contours for field 
strength.}
\label{fig:allss}
\end{figure}

Dense cores formed under different conditions may have different degrees
of magnetization. In our formulation of the problem, the degree of 
magnetization is controlled by the parameter $H_0$. To gauge its effects
on the collapse solution, we consider three cases, with $H_0=0.125$, 
$0.25$ and $0.5$ ($\lambda = 8.38$, 4.51, and 2.66, respectively).
The results are plotted in similarity coordinates and reduced flow
variables in Fig.~\ref{fig:allss},
with the non-magnetic case $H_0=0$ also shown for reference.
The solutions in all 
three magnetized cases look qualitatively similar. The most prominent
feature in each case is the dense, flattened structure in the equatorial
region -- a pseudodisk. Another important feature is the magnetically 
dominated, low-density region around the axis -- a polar cavity, where 
matter drains preferentially along the field lines onto the pseudodisk.
As $H_0$ increases, the magnetic field dominates a larger region near
the polar axis, as shown by the constant-$\beta$ contours (particularly 
the $\beta=1$ contour). As a result, the size of the pseudodisk increases 
as well (at any given time), for the obvious reason that the matter in
the cavity has been emptied onto the disk. A more subtle effect 
is on the mass accretion rate onto the center point mass, as measured
by the reduced mass $m_0$ defined in equations~(\ref{centralmass}) and 
(\ref{Mdotin}). 
We find that the values of $m_0$ in all four cases are within $2\%$ 
of $0.975(1+H_0)$.
 
The $1+H_0$ dependence arises because it represents the
overdensity factor of the initial 
configuration through partial support by magnetic fields.
As found also by Galli \& Shu (1993b), other effects,
such as the larger outward speed for signal propagation and
the smaller inward speed of actual infall because of magnetic retardation, 
largely cancel in their net effect.  There is a tendency for the more 
strongly magnetized case to have a slightly lower coefficient in front 
of the factor $1+H_0$. The exact reason for the slight decline of 
the computed numerical coefficient with $H_0$ is unclear. Semi-analytical 
theory suggests that the coefficient should actually have increased 
slightly to 1.05 for $H_0 \gg 1$ (Li \& Shu 1997).
The Alfv{\'e}n limiter may play a role by its artificial depression
of the outward signal speed that initiates the infall.  But small effects
of numerical origin (particularly near the
origin where the magnetic field is very strong and numerical diffusion 
is a concern; see next paragraph) cannot be ruled out. In any case, the difference in the numerical coefficient is small,
and we believe that the expression of equation (\ref{magMdotin}) 
provides a good practical approximation at the 5\% level
for the full range of $H_0$ from 0 to $\infty$. 

We were not able to obtain converged collapse solutions for more strongly
magnetized cores with $H_0$ significantly greater than 0.5. Part of
the difficulty appears in the form of a
high magnetic barrier created by a split monopole as mass and field
is accreted into the origin (see Li \& Shu 1997).
As time advances, the barrier becomes larger relative to numerically resolved
cells near the center.  In the 
ideal MHD limit, field may never leave the origin and the flow will always 
be smooth.  But, if there were ambipolar diffusion, one would expect field 
to escape from the origin into the pseudodisk, where it would expand 
against the already magnetized fluid waiting to be accreted, slowing the 
accretion rate until enough mass piles up to overwhelm the effect of the 
excess field \footnote{Without axial symmetry, the escaping field lines 
and incoming fluid could perhaps slip past each other, avoiding this 
oscillatory process}.  Then accretion would proceed at an accelerated 
rate, until again, field leaks out of the origin and the process repeats, 
ever growing in amplitude.  This process is mimicked to some extent by 
the unavoidable numerical diffusivity of the MHD code.  For this reason, 
the code cannot run forever and eventually halts due to growing 
oscillations.  As the pivotal state magnetic field strength increases, 
so does this effect.  Nevertheless, accretion is relatively constant.
\footnote{There appears to be a slight decrease in central accretion 
over the course of the simulation (see Fig.~3d for an example). We 
believe this may arise because of the small diffusive effect of field 
lines slipping outward from the central cell, an effect exacerbated 
by the artificial viscosity used in Zeus2D to mediate (accretion) shocks.}
The central accretion rate increases with increasing $H_0$, 
approximately in proportion to $(1+H_0)$.

\section{Discussion}

\subsection{Protostellar Mass Accretion Rate}

\begin{figure}
    \centering
    \leavevmode
\epsfig{file=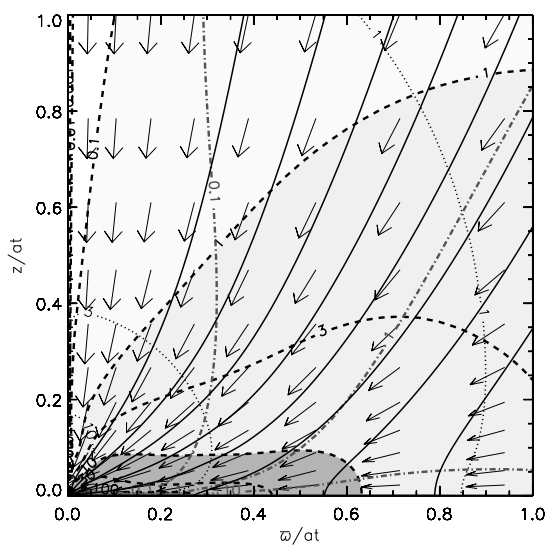,width=.48\textwidth,clip=}
\caption{Collapse solution of the $H_0=0.25$ case with an initial inward radial velocity of $0.5a$.
The contours of constant 
self-similar density $4\pi Gt^2\rho$ are plotted as dashed lines, 
with the shades highlighting the high density regions. The magnetic 
field lines are plotted as solid lines,  with contours of constant 
$\beta$ (dash-dot-dashed) superposed. The velocity in every fifth 
cell is shown by unit vectors, with its magnitude in units of the 
sound speed given by the dotted contours.}
\label{fig:inwardmotion}
\end{figure}

An interesting question is how much our computed mass-accretion rates
would have changed if we had allowed for the presence of inward motions in
the pivotal state.  We have performed a number of numerical simulations
starting with the density and magnetic field configuration of equilibrium toroids
(characterized non-dimensionally by the over-density parameter $H_0$), and added an
initially uniform, radially inwardly directed, velocity field of $0.5 \, a$.
An example is shown in Figure~\ref{fig:inwardmotion}.  As we commented upon at the beginning, the
results retain their exact self-similarity in spacetime.  The qualitative features of
the calculation -- the formation of a pseudodisk, the trapping of the magnetic flux
by the central cell to yield a split-monopole geometry, etc. -- also remain unchanged from
the corresponding case starting the gravitational collapse from rest. The only
substantive modification comes from an increase by about a factor of 2 in the mass-accretion
rate onto the central object.

Since unstable cloud cores probably have dimensionless
mass-to-flux ratios $\lambda = $ 1--3, whereas
T Tauri stars have $\lambda >$ 5000, the assumption of
field freezing must break down at high collapse densities
(Nakano \& Umebayashi 1980, Li \& Shu 1997).
Whether flux loss in forming protostars ultimately
occurs by magnetic reconnection (e.g., Mestel \& Strittmatter
1967, Galli \& Shu 1993b) or by C-shocks (Li \& McKee 1996, Li 1998a,
Ciolek \& K\"onigl 1998) remains to be seen.  We elaborate on these
possibilities in our companion paper after we include the effects
of rotation.  Self-similar
collapse calculations carried out with the assumption of
strict field freezing yield important ``outer solutions''
for more realistic calculations of the flux-loss problem
in the inner regions (e.g., Li 1998b).

An interesting numerical instability inevitably develops to complicate all of our actual calculations.
A point mass acting as a split magnetic monopole grows linearly with time at the origin.  
But, accretion into the point mass occurs primarily through the pseudodisk.
Gas in the pdesudodisk feels a magnetic barrier from the split monopole field
that opposes rapid accretion.  As the split monopole field grows,
mass may pile up in the pseudodisk until it becomes too heavy for the field to support.
Until this time, there is lower than average accretion into the origin.
Then the excess mass rushes into the point mass,
dragging more field to be assimilated by the split monopole.
At this time, there is above average accretion into the origin.
Thus, the magnetic barrier grows, and mass must pile up in the pseudodisk
for a longer time interval, before it can join the point mass,
and the oscillations in accretion rate grow in amplitude.
The instability has a purely numerical origin 
because the strictly self-similar, ideal problem contains,
of course, no dimensional capability for supporting a temporal period.
Nevertheless, the interesting question remains whether the situation
with more realistic physics, mimicked by the numerical effects described here,
might not sustain relaxation oscillations of a type that are 
responsible for FU Orionis outbursts.

\subsection{Limited Role for Magnetic Tension}

There is another important lesson to be learned from the
self-similar collapse calculations.
Contrary to earlier speculations made in
the literature, magnetic tension in the envelope
never suffices to suspend the envelope
against the gravity of the growing protostar.  
 \footnote{The concept that magnetic tension can stop infall implicitly 
assumes that field lines at large distances are held laterally apart 
by external agencies.  Our example avoids this assumption by allowing 
the matter and field to have dynamics all the way to infinity.  
Depending on the nature of the applied outer boundary conditions, 
caveats may need to be added if the outer parts of a cloud are 
subcritical (see Shu, Li \& Allen 2003 for a detailed analysis
of this problem).} When the wave
of infall reaches successively larger portions of
the envelope, those portions move inward
to fill the hole and becomes part of the infalling material.
In the specific case of the magnetized singular isothermal
toroid, without some other influence,
the mass infall rate into the central regions
continues indefinitely at the constant rate given approximately by
equation (\ref{magMdotin}).  In
all cases, the inclusion of the effects of
magnetic forces (when $H_0 > 0$) {\it increases} the rate of
central mass accumulation in comparison with
the nonmagnetized case ($m_0 = 0.975$), and does not decrease it.
The main reason for the increase is that the cloud core
is denser on average before it begins to collapse
when supported by magnetic fields than when
this support is absent.  The slowing down of
the infall speed by magnetic forces once collapse
begins is largely offset by the increase of the signal
speed at which the wave of infall propagates outward
(Galli \& Shu 1993a,b). 

\subsection{Defining Stellar Masses}

Because magnetic tension in the envelope is unable to halt
the continued accretion, one of three things
must happen if stellar masses are to result from the
infall process (Shu et al. 1999).  First, the growing star might
run out of matter to accumulate.  For example,
the wave of infall might run into a vacuum because
the material previously there has already fallen into
some other star.  This alternative requires
the star formation process to be nearly 100\% efficient,
which we know is not the case except, possibly, in
the relatively rare cases of bound
cluster formation. 

Second, the star might move out of its surrounding
cloud core.  Large motions relative to the local pit in the
gravitational potential are likely only in very crowded
regions of star formation, which we have decided to exempt from
the present discussion (but see Shu et al. 1999).

Third, the material otherwise destined
to fall onto the star might be removed
by some hydrodynamic, magnetohydrodynamic, or
photo-evaporative process.  Because of the ubiquity of bipolar
outflows in star-forming regions, Shu \& Terebey (1984) implicated
YSO winds as the dominant practical mechanism by which forming stars
help to define their own masses, particularly in
isolated regions of low-mass star-formation.  In a 
future study, we will introduce a protostellar wind into the collapse 
simulation, aiming to determine not only its effects on the mass
distribution, but also the kinematics of the wind-swept ambient 
materials.

\subsection{Flattened Mass Distribution}

\begin{figure}
    \centering
    \leavevmode
        \epsfxsize=.48\textwidth \epsfbox{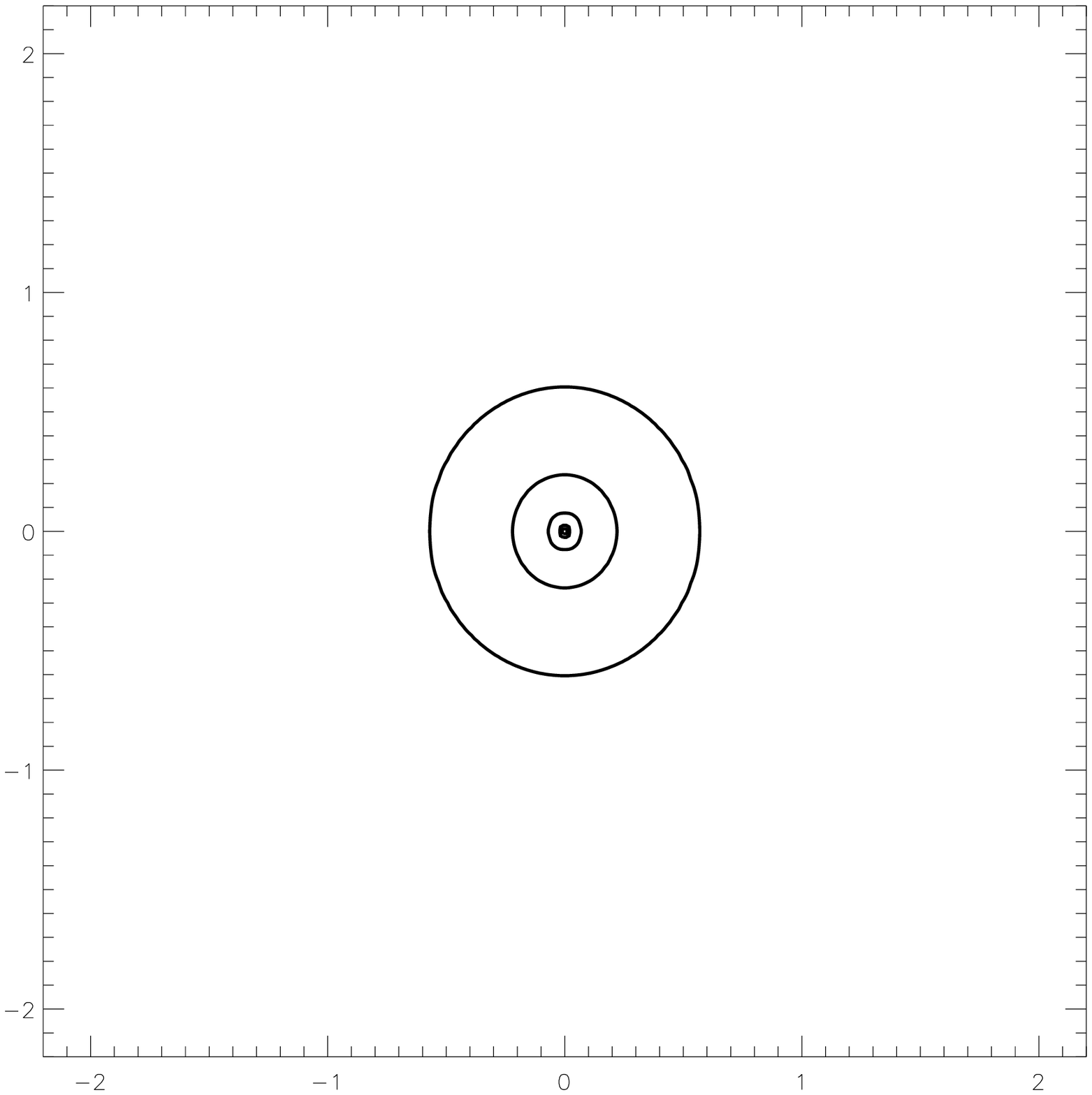} 
        \epsfxsize=.48\textwidth \epsfbox{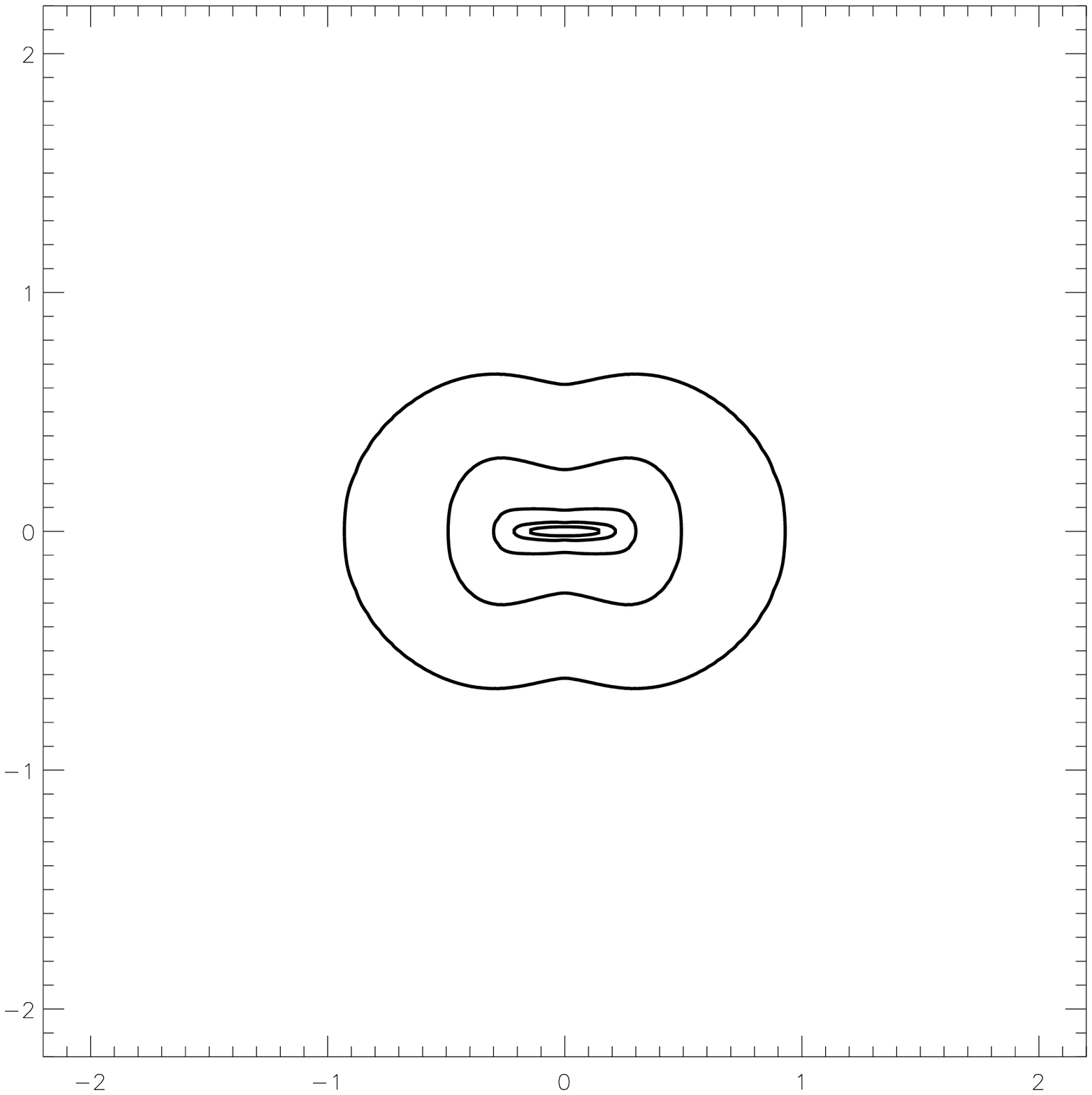} 
        \epsfxsize=.48\textwidth \epsfbox{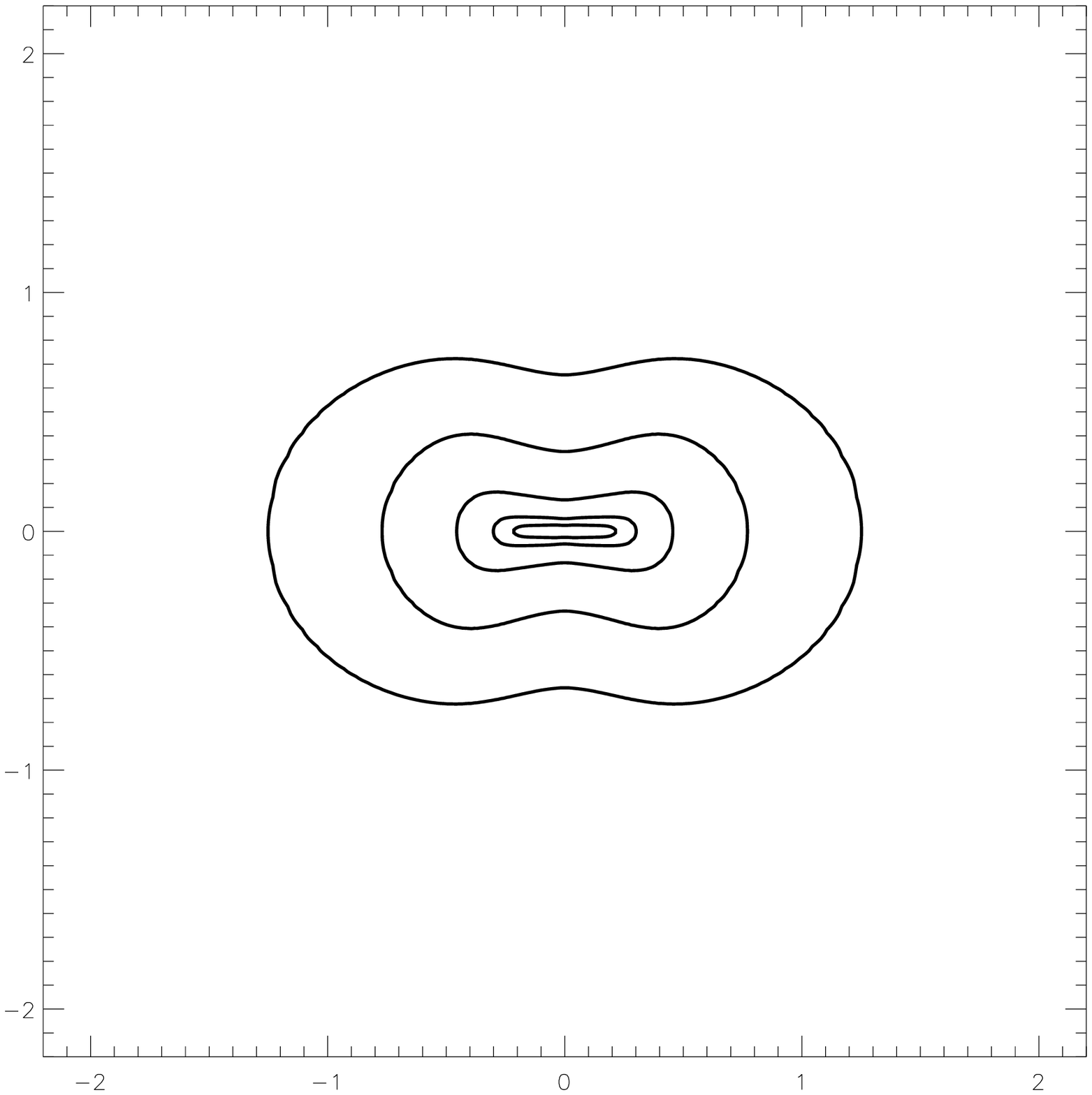} 
        \epsfxsize=.48\textwidth \epsfbox{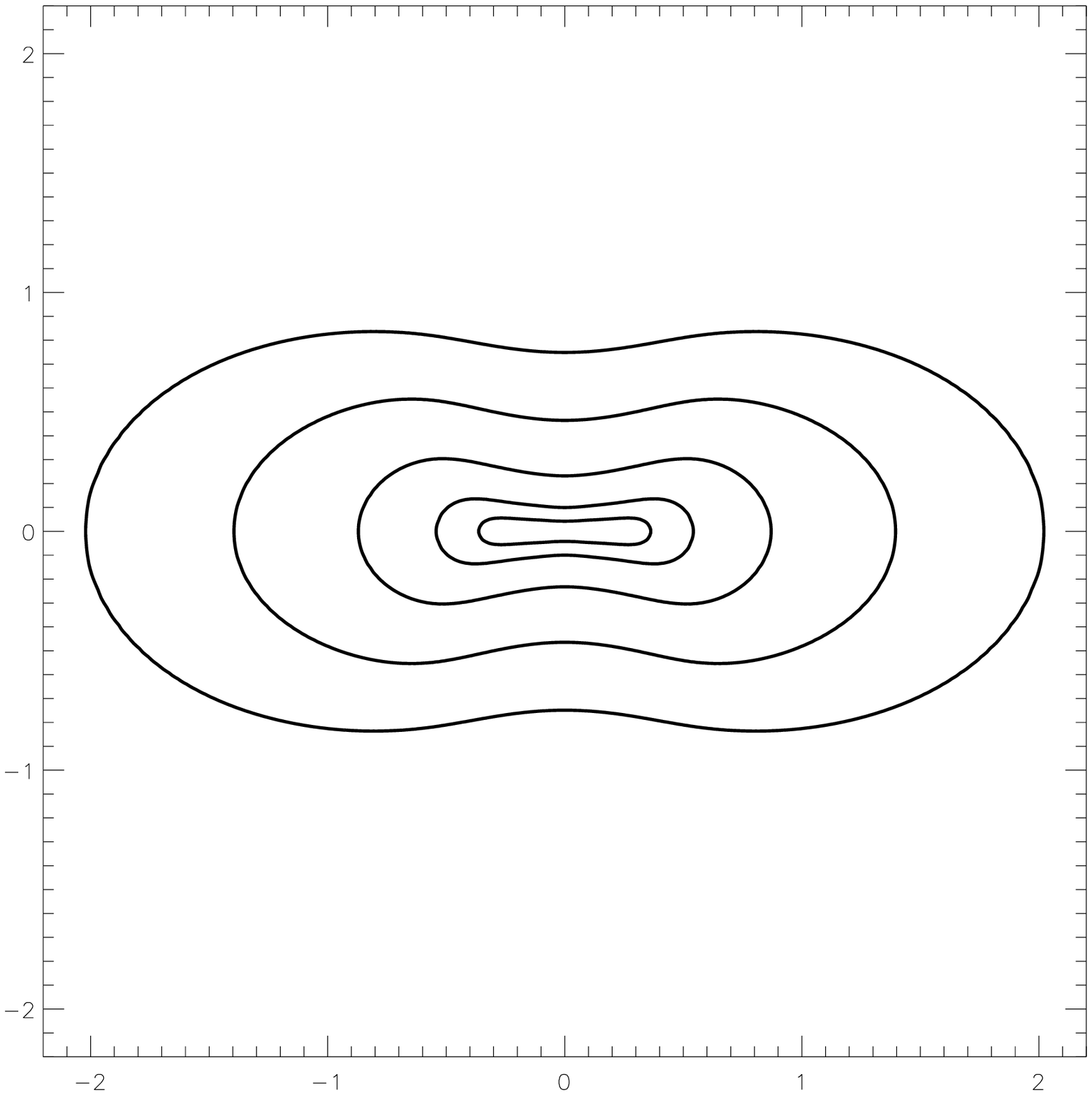}
\caption{Logarithmically spaced iso-mass column-density contours (viewed 
equatorially) for collapse solutions with $H_0=0,
0.125, 0.25$ and $0.5$, showing the increasing degree of elongation
on small scales for the magnetized cases. Mass columns are integrated for 
a cloud extending to $x=\pm 10$.  From outward to inward, contours are 
at $N=7.4,11,16,25,$ and $36$.  To convert to physical units, multiply 
the axis by $at$ and mass column-density by $a/4\pi G t$.  For example, 
for $a=0.2$~km~s$^{-1}$ at $t=10^5$~yr, 1 unit increments 
in $x$ correspond to $6\times 10^{16}$~cm and the outermost 
(innermost) mass column-density contour
is at $0.056$~g~cm$^{-2}$ ($0.27$~g~cm$^{-2}$). }
\label{fig:column}
\end{figure}

Using a simple thin-shell model, Li \& Shu (1996) computed the kinematic
signature of an X-wind propagating into a static, magnetized SIT
and showed that the mass-with-velocity distribution was compatible
with observations (for a summary, see Shu et al. 2000).  However, a truly
self-consistent model would account for the dynamic infall in the
inner regions of the molecular cloud core that produced the central 
source. The anisotropy in the mass distribution of an initially magnetically
supported structure 
is amplified during the collapse, creating a more extended low-density
polar region and a denser equatorial pseudodisk. 
This effect can be see in Fig.~\ref{fig:column}, where
we plot the iso-column density contours viewed perpendicular to 
the axis, for the cases of $H_0=0, 0.125, 0.25$ and $0.5$ shown in 
Fig.~\ref{fig:allss}.   Note that the column density contours become 
more elongated closer to the protostar, reflecting the formation 
of pseudodisk in the inner collapse region. It remains to be seen what
are the modifications to be introduced in the calculations of Li \& Shu 
(1996) by these effects, particularly in young outflow sources before 
the swept-up lobes have had a chance to penetrate far beyond the
wave of infall.

\subsection{Effect of Rotation}

YSO outflows arise, many people believe, because of a fundamental
interaction between the rapid rotation in a Keplerian disk and some
strongly magnetized object (either the central star or the disk itself).
By ignoring the effects of rotation in this paper, we have chosen
to focus on the pure effects of interstellar magnetic 
fields on the dynamics of gravitational collapse. However, as discussed above, magnetic fields by themselves do not resolve the fundamental question of
why stellar masses result as a product of the gravitational collapse of
molecular clouds.  Fortunately, rotation 
is observed ubiquitously in star-forming dense cores
(e.g., Goodman et al. 1993). 
Although typically not fast enough to contribute significantly to 
the core support, it is presumably responsible for the formation of Keplerian
disks around young stars.  In many studies, disk sizes are computed
under the assumption that the total angular momentum observed in the cores
is entirely carried into the collapsed object, although there
may be some viscous redistribution within the Keplerian accretion disk.
The following paper in this series, which finds relatively
strong magnetic braking to occur even during the collapse process,
casts doubt on the validity on this basic assumption when molecular 
cloud cores
are realistically rotating {\it and} magnetized.

\acknowledgments{Support for this work was provided in Taiwan in part
by grants from
Academia Sinica and the National Science Council, and in the United
States by grants from the
National Science Foundation and NASA.}

\clearpage

\end{document}